\documentclass{article}
\usepackage{hyperref}
\usepackage{graphicx}
\usepackage{amsmath,amssymb}
\usepackage{color}

\newcommand{\R}{\mathbb{R}}
\DeclareMathOperator*{\argmax}{arg\,max}

\title{Evolution of truncated and bent gravity wave solitons: the Mach
  expansion problem}

\author{Samuel Ryskamp$^1$,
  Michelle D.~Maiden$^1$, Gino Biondini$^2$, Mark
  A.~Hoefer$^1$\thanks{Email address for correspondence:  hoefer@colorado.edu}}

\date{\normalsize
$^1$Department of Applied Mathematics, University 
of Colorado, Boulder, CO 80309, USA,\\
$^2$Department of Mathematics, State University of New York,
  Buffalo, NY, USA}

\begin{document}

\maketitle

\begin{abstract}

  The dynamics of initially truncated and bent line solitons for the
  Kadomtsev-Petviashvili (KPII) equation modelling internal and
  surface gravity waves are analysed using modulation theory.  In
  contrast to previous studies on obliquely interacting solitons that
  develop from acute incidence angles, this work focuses on initial
  value problems for the obtuse incidence of two or three partial line
  solitons, which propagate away from one another.  Despite
  counterpropagation, significant residual soliton interactions are
  observed with novel physical consequences.  The initial value
  problem for a truncated line soliton---describing the emergence of a
  quasi-one-dimensional soliton from a wide channel---is shown to be
  related to the interaction of oblique solitons.  Analytical
  descriptions for the development of weak and strong interactions are
  obtained in terms of interacting simple wave solutions of modulation
  equations for the local soliton amplitude and slope.  In the weak
  interaction case, the long-time evolution of truncated and large
  obtuse angle solitons exhibits a decaying, parabolic wave profile
  with temporally increasing focal length that asymptotes to a
  cylindrical Korteweg-de Vries soliton.  In contrast, the strong
  interaction case of slightly obtuse interacting solitons evolves
  into a steady, one-dimensional line soliton with amplitude reduced
  by an amount proportional to the incidence slope.  This strong
  interaction is identified with the ``Mach expansion'' of a soliton
  with an expansive corner, contrasting with the well-known Mach
  reflection of a soliton with a compressive corner.  Interestingly,
  the critical angles for Mach expansion and reflection are the
  same. Numerical simulations of the KPII equation quantitatively
  support the analytical findings.

\end{abstract}

\section{Introduction}
\label{sec:introduction}

The oblique interaction of solitary waves or solitons is a fundamental
problem in fluid dynamics and nonlinear sciences more broadly.  Early
theoretical consideration of this problem for acute, collisional
angles of incidence dates back to
\cite{miles_obliquely_1977,miles_resonantly_1977} where weak and
strong gravity water wave soliton interactions were shown to be
dependent upon the incidence angle and soliton amplitudes.  In the
case of weakly interacting oblique solitons, a sufficiently small
incidence angle leads to the approximate linear superposition of the two
solitons accompanied by a phase shift.  The strong interaction case at
large acute angles leads to a resonant triad of obliquely interacting
solitary waves, the so-called Miles resonant soliton.  Miles used his
theory to identify the long-time dynamics of regular and Mach
reflection of a soliton incident upon a compressive corner or wedge.

\begin{figure}
  \centering
  \includegraphics[scale=.25]{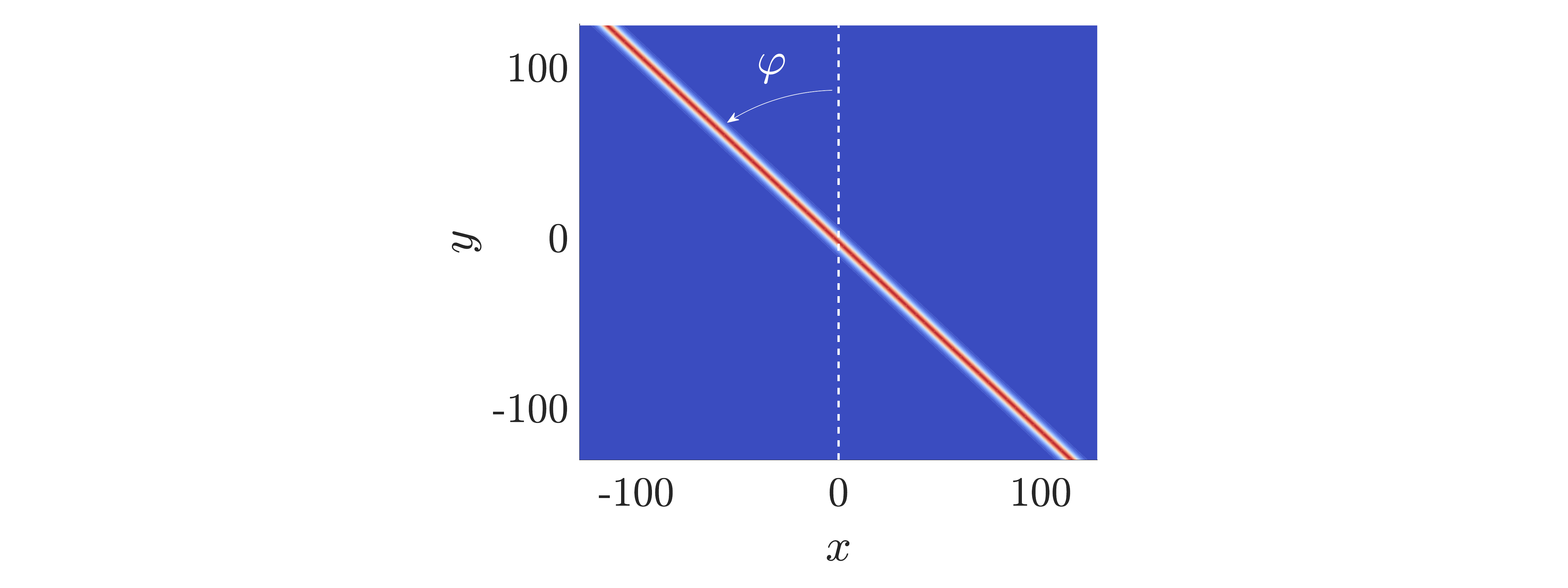}
  \caption{Contour plot of a line soliton solution and its slope
    parametrisation.  The soliton propagation slope is $q = \tan
    \varphi$.}
  \label{fig:line_soliton}
\end{figure}
Building upon original water wave experiments by
\cite{perroud_solitary_1957}, Miles' studies have since been expanded
and refined with the help of laboratory experiment
\cite{melville_mach_1980,li_mach_2011,kodama_kp_2016} and field
observations \cite{wang_oblique_2012,ablowitz_nonlinear_2012},
numerical simulation
\cite{funakoshi_reflection_1980,tanaka_mach_1993,porubov_formation_2005,tsuji_oblique_2007,biondini_soliton_2009,kodama_soliton_2009,li_mach_2011,kao_numerical_2012,chakravarty_numerics_2017}
and exact $N$-soliton solutions
\cite{kodama_young_2004,biondini_line_2007,biondini_soliton_2009,chakravarty_soliton_2009,kodama_kp_2010,kodama_kp_2016}
of the Kadomtsev-Petviashvili (KP) equation and its higher order
generalisations.  The KP equation is a generic model of weakly
nonlinear, weakly dispersive unidirectional waves with weak transverse
variation \cite{kadomtsev_stability_1970}
\begin{equation}
  \label{eq:KP}
  \begin{split}
    (u_t + uu_x + u_{xxx})_x + u_{yy} = 0, \quad (x,y) \in
    \mathbb{R}^2,
    \quad t > 0,
  \end{split}
\end{equation}
originally derived in the context of shallow water waves by
\cite{ablowitz_evolution_1979} and for internal waves by
\cite{grimshaw_evolution_1981}.  In applications such as shallow water
waves, the nondimensionalisation is achieved by entering the reference
frame moving with the long wave speed $\sqrt{gh}$ ($g$ is
gravitational acceleration and $h$ is the mean fluid depth), scaling
time by $\tau$, scaling the longitudinal and transverse lengths by the
typical wavelengths $\lambda_x$ and $\lambda_y$, respectively, and
scaling the surface disturbance height by the typical amplitude
$\eta$.  The KP equation arises when weak nonlinearity, long wave
dispersion, and weak transverse variation balance according to,
respectively,
$\eta/h \sim (h/\lambda_x)^2 \sim (\lambda_x/\lambda_y)^2 \ll 1$.  The
particular scaling \eqref{eq:KP} manifests if we set
$\tau = \lambda_x^3/(h^2\sqrt{gh})$,
$\lambda_y = \lambda_x^2/(\sqrt{2}h)$, and
$\eta = 2 h^3/(3\lambda_x^2)$ for some $\lambda_x \gg h$.  The
asymptotic validity of the KP equation requires small transverse wave
curvature for quantitative comparison with physical observations.
Geometric and higher order asymptotic considerations can be used to
achieve even better agreement between theory and experiment
\cite{kodama_kp_2016}.

This version of the Kadomtsev-Petviashvili equation is known as the
KPII equation---the KPI equation occurs when $+u_{yy} \to -u_{yy}$.
The KPII equation is a completely integrable equation
\cite{ablowitz_solitons_1991} that admits a two-parameter family of
stable line soliton solutions
\begin{equation}
  \label{eq:KPsoliton}
  u(x,y,t) = a \,
  \mathrm{sech}^2\left(\sqrt{\frac{a}{12}}(x + q y - ct)\right), \quad
  c=\frac{a}{3}+q^2 ,
\end{equation}
uniquely determined by the amplitude $a > 0$ and soliton inclination
from the $y$-axis or slope $q \in \R$.  See
Fig.~\ref{fig:line_soliton} for a representative example.  We note
that the weak transverse scaling used in the derivation of the KP
equation \eqref{eq:KP} incorporates this assumption so that the slope
$q$ can be an order one quantity.

\begin{figure}
  \centering
  \includegraphics[scale=.25,trim={0 9cm 0 0}]{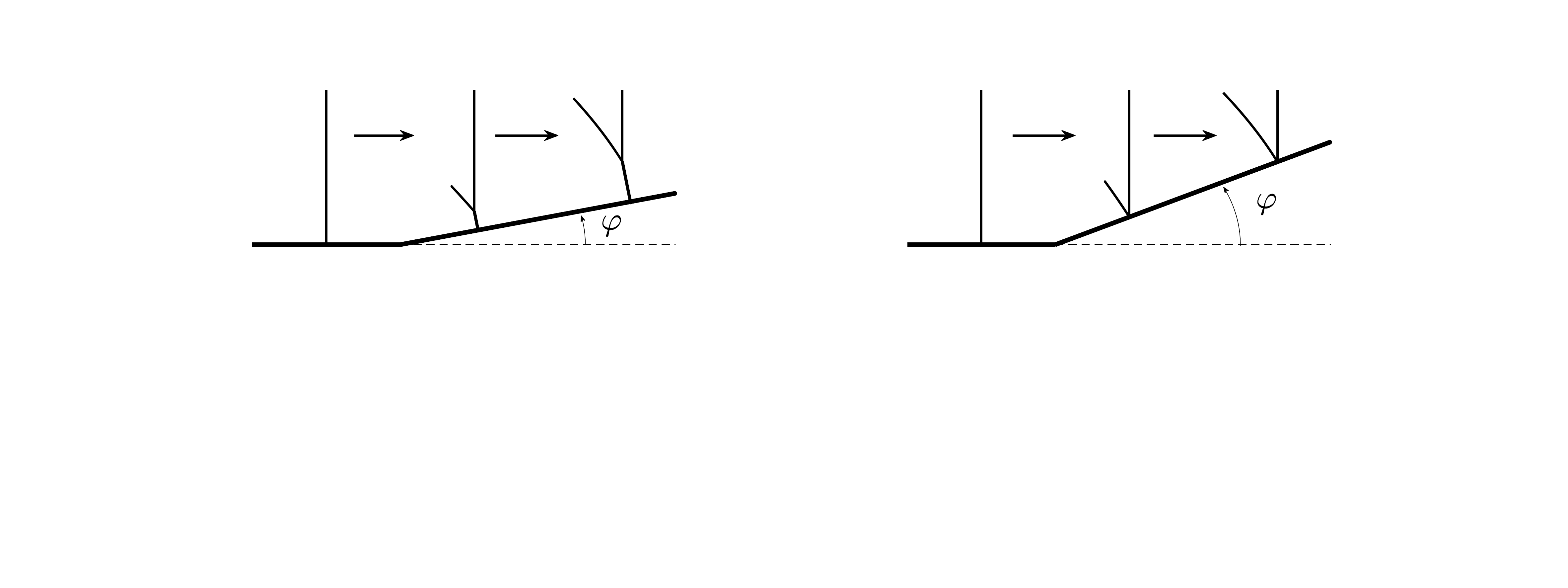}
  \caption{Schematic of Mach (left) and regular (right) reflection of
    a soliton impinging upon a compressive corner.}
\label{fig:mach_reflection}
\end{figure}

The Mach reflection problem has an interesting history that begins
with Ernst Mach's research on shock wave interactions in gas dynamics.
Utilising two separated electric spark sources between two glass
plates, one covered in soot, \cite{mach_uber_1875} generated
cylindrical shock fronts that left residual patterns from their
interaction.  From these experiments, Mach keenly discerned two types
of shock interaction---termed regular and irregular---that depended
upon the interacting fronts' obliqueness.  Both cases involved a
reflected shock but the irregular interaction for sufficient
obliqueness resulted in a triple point where three regions of
different pressures and densities met.  An additional wave was
generated from the triple point and has become known as the Mach stem
resulting from Mach reflection of shock waves
\cite{krehl_discovery_1991}.  Motivated by the so-called hydraulic
analogy between two-dimensional supersonic gas dynamics and
supercritical shallow water waves, \cite{gilmore_analogy_1950}
reconsidered regular and Mach reflection of shocks by obliquely
interacting two hydraulic jumps.  This phenomenon was also observed in
shallow ocean waves \cite{cornish_waves_1910}. But the theoretical
interpretation of Mach reflection in water waves had to wait until the
seminal work of \cite{miles_obliquely_1977,miles_resonantly_1977}, in
which two obliquely interacting solitary waves were described. The
irregular or Mach reflection case is embodied in the resonant or
Y-shaped solitary wave solution.  Regular reflection is described by
the X-shaped solitary wave solution.  Regular and Mach reflection can
also be formulated as an initial-boundary value problem in which a
semi-infinite shock or soliton propagating parallel to a wall impinges
upon a wedge or corner.  Depending upon the corner angle, the ensuing
dynamics lead to the spontaneous generation of a reflected wave and,
in the case of irregular or Mach reflection, the additional generation
of a Mach or stem wave in which a resonant triad of three waves meet
and propagate away from the wall.  See Fig.~\ref{fig:mach_reflection}
for a schematic of the two reflection types.  Because the KP equation
\eqref{eq:KP} is a generic, universal model of weakly nonlinear,
dispersive, two-dimensional wave patterns, regular and Mach reflection
are fundamental to the description of multidimensional nonlinear
waves.

One approach to describe Mach reflection of solitons is the use of
exact solutions of the KPII equation \cite{li_mach_2011}.  A
classification of 2-soliton solutions in
\cite{kodama_young_2004,biondini_line_2007,chakravarty_soliton_2009}
was used to identify two particular 2-soliton solutions
whose parameters can be chosen to satisfy the requisite structure of
regular and Mach reflected waves.  To describe the soliton-corner
initial, boundary value problem, a nonlinear method of images is
applied and hypothesised to locally describe the long-time dynamics.
This results in the critical angle $\varphi_{\rm cr}$---corner
inclination measured from the positive $x$-axis---for the transition
from regular, $\varphi > \varphi_{\rm cr}$, to Mach, $0 < \varphi <
\varphi_{\rm cr}$, reflection of an incident soliton with amplitude
$a$ as
\begin{equation}
  \label{eq:64}
  \tan{\varphi_{\rm cr}} = \sqrt{a}.
\end{equation}
Numerical simulations of initially V-shaped waves were used to justify
the long-time, locally 2-soliton solution hypothesis for the
soliton-corner initial, boundary value problem
\cite{kodama_soliton_2009,kao_numerical_2012}.  Herein lies a subtle
difference between oblique soliton interaction---described by exact
2-soliton solutions---and a soliton incident upon a corner, which
involves transient dynamics that, after long enough times, are locally
described by 2-soliton solutions.

\begin{figure}
  \centering
  \includegraphics[scale=.25,trim={0 5cm 0 0},clip]{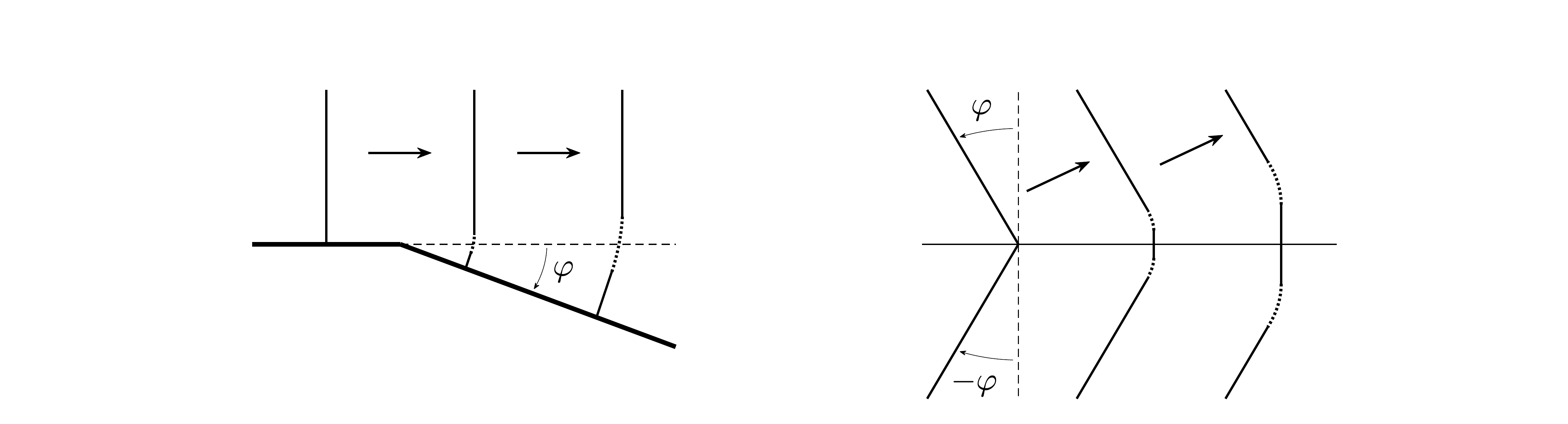}
  \includegraphics[scale=.25,trim={0 5cm 0 0},clip]{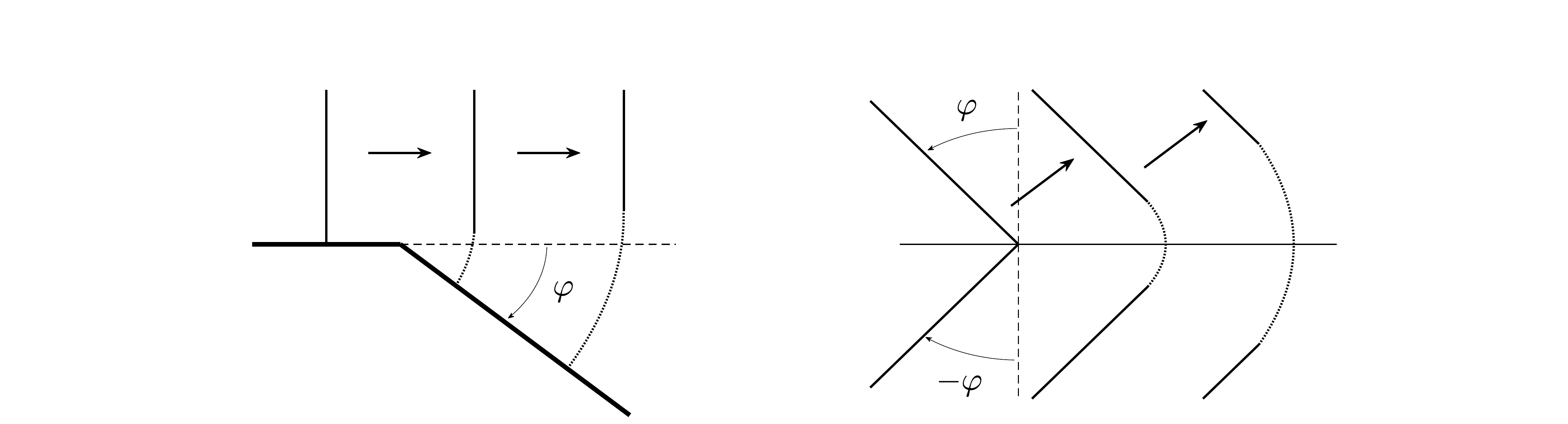}
  \caption{Left: diagram of Mach expansion for a front encountering a
    reverse wedge boundary with small $\varphi$ (top) and large
    $\varphi$ (bottom). Note the emergence of a new, smaller amplitude
    line soliton along the boundary when $\varphi$ is small.  Right:
    bent soliton initial conditions are identical for the purposes of
    analysis.}
  \label{fig:mach_expansion}
\end{figure}

The aforementioned regular/Mach reflection problem involves a
\textit{compressive} corner with angle $\varphi$ measured
counterclockwise from the positive $x$-axis.  In this paper, we
consider the problem of a soliton incident upon an \textit{expansive}
corner, opening in the opposite direction so that $\varphi$ is
measured clockwise from the positive $x$-axis.  This problem is rather
different from the regular/Mach reflection problem in many respects
but we find an interesting parallel.  The critical corner angle that
separates regular expansion and Mach expansion for an incident soliton
with amplitude $a$ occurs precisely at $\varphi_{\rm cr}$, the same
critical angle separating regular and Mach reflection in Equation
(\ref{eq:64}).  Regular expansion occurs when
$\varphi > \varphi_{\rm cr}$ and leads to the development of a
decaying parabolic wave that connects the incident soliton to the
wall.  Mach expansion when $0 < \varphi < \varphi_{\rm cr}$ involves
the development of a new soliton perpendicular to the wall with
reduced amplitude relative to the incident soliton.  The development
of this soliton is the expansion analogue of the Mach stem in the
reflection case.  See Fig.~\ref{fig:mach_expansion} for a schematic of
the two cases.

A new approach is required to describe regular and Mach expansion of
solitons because the dynamics are not described by multisoliton
solutions.  While the transient dynamics of Mach reflection are
subtle, crucial aspects of Mach expansion are transient and are not
steady in a local reference frame.  By making a slowly varying
assumption, we approximately describe the full expansion dynamics
developing from initial value problems by appropriate modulation of
the line soliton (\ref{eq:KPsoliton}).  Our KPII numerical simulations
demonstrate that modulation theory is effective at uncovering key,
quantitative features of the nonlinear wave dynamics.

\begin{figure}
  \centering
  \includegraphics{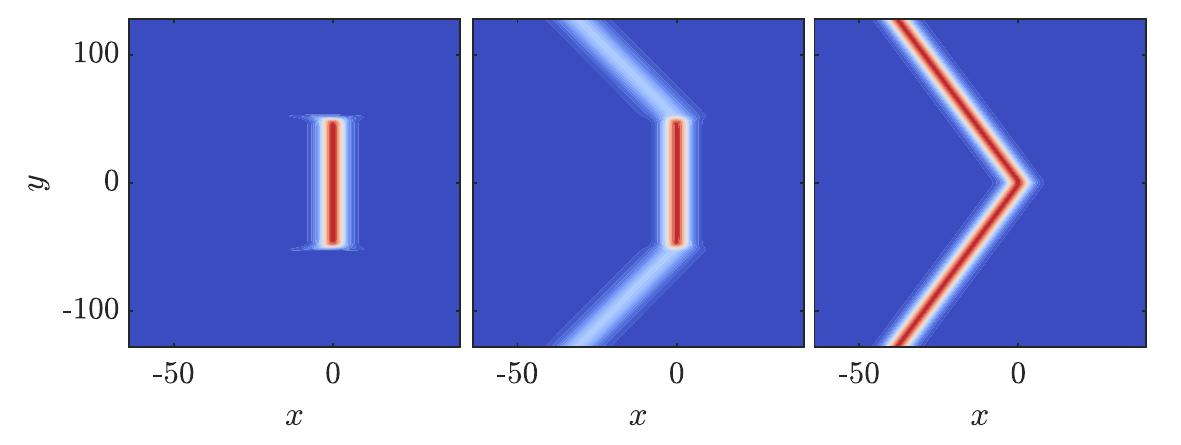}
  \caption{Initial data corresponding to a truncated line soliton
    (left), a bent-stem soliton (centre), and a bent soliton (right).}
  \label{fig:initial_data}
\end{figure}

In order to describe the soliton-corner expansion problem, we take a
seemingly circuitous route by considering three classes of initial
value problems, depicted in Fig.~\ref{fig:initial_data}, for the KPII
equation \eqref{eq:KP}.  From left to right, we identify the data as
truncated, bent-stem, and bent soliton initial conditions.  The bent
soliton case corresponds to the appropriate reflection, via the
nonlinear method of images \cite{kodama_kp_2016}, needed to describe
regular/Mach expansion as depicted in Fig.~\ref{fig:mach_expansion}.
Bent initial conditions are equivalent to a single soliton moving
along a boundary, where the boundary suddenly angles away from the
front.

The reason for considering these three classes of initial data in turn
is both mathematical and physical.  From a mathematical point of view,
each initial condition limits to the next so that their solution is
informed by the previous and they share some solution properties.
Moreover, these are among the simplest and more natural kinds of
non-solitonic initial conditions for the KP equation one could
consider.  Although the data consists of modulated solitons, it does
not contain exact soliton solutions.  In contrast to our approach, the
primary analytical means by which all previous studies have
interpreted related non-solitonic initial value problems is by
approximating the evolution of initial configurations with exact
$N$-soliton solutions of the KP equation.  See, for example,
\cite{biondini_soliton_2009,kao_numerical_2012,chakravarty_numerics_2017}.

From a physical point of view, the truncated soliton initial data
models the emergence of a quasi-one-dimensional soliton from
transverse confinement such as the internal ocean solitons generated
at the front of a river plume
\cite{pan_analyses_2007,wang_internal_2017} and a surface wave
soliton when a channel suddenly widens.  The rapid deceleration or
transcritical propagation of a ship in open, shallow water can
similarly launch a bent-stem or bent soliton from the ship's prow
\cite{li_three-dimensional_2002}, dependent upon the prow shape.
Finally, internal wave solitons are ubiquitous in the world's oceans
\cite{jackson_atlas_2004,wang_oblique_2012} and topography
significantly impacts their propagation and interaction
\cite{yuan_topographic_2018}.  The classes of initial data in
Fig.~\ref{fig:initial_data} represent cases where the modulated
solitons propagate away from one another.  Nevertheless, their
interaction is significant.

We study these initial value problems using modulation theory in which
the local soliton amplitude $a = a(y,t)$ and slope $q = q(y,t)$ are
allowed to vary slowly in space and time.  There are several
approaches to derive the effective modulation equations.  In Appendix
\ref{sec:deriv-modul-equat}, we provide a derivation using multiple
scale perturbation theory of the equations
\begin{subequations}
  \label{eq:modulationEqs}
  \begin{align}
    \label{eq:25}
    a_t + 2 q a_y + \frac{4}{3} a q_y &= 0,  \\
    \label{eq:27}
    q_t + 2 q q_y + \frac{1}{3} a_y &= 0 .
  \end{align}
\end{subequations}
The dynamical equation (\ref{eq:25}) for the amplitude results from an
appropriate orthogonality condition and the slope equation
(\ref{eq:27}) results from a consistency condition of the modulated
phase.  The modulation equations \eqref{eq:modulationEqs} were also
derived from a variable coefficient KP equation in
\cite{lee_upstreamadvancing_1990} and by an averaged Lagrangian
approach in \cite{neu_singular_2015,grava_numerical_2018}.  They are
also a limiting case of the more general KP-Whitham modulation
equations for periodic waves \cite{ablowitz_whitham_2017} in the case
of $x$ independent modulations of a line soliton
\cite{biondini_integrability_2019}.  These soliton modulation
equations are equivalent to the equations modelling the isentropic flow
of a polytropic gas with density $\propto a^{3/2}$, velocity $\propto
q$, and ratio of specific heats $\gamma = 5/3$.  The linearisation of
the modulation equations \eqref{eq:modulationEqs} for small $|q|$ was
used in Kadomtsev and Petviashvili's original paper to determine the
stability of line soliton solutions to the KP equation
\cite{kadomtsev_stability_1970}.

Despite variation in only one spatial dimension, the corresponding
modulated line solitons exhibit non-trivial two-dimensional structure.
In particular, once a modulation solution $a(y,t)$, $q(y,t)$ is
obtained, the modulated soliton is reconstructed by projection onto
(\ref{eq:KPsoliton}) according to
\begin{equation}
  \label{eq:39}
  \begin{split}
    u(x,y,t) &\sim a(y,t) \,\mathrm{sech}^2\left (
      \sqrt{\frac{a(y,t)}{12}} \xi \right ), \\
    \xi &= x + \int_0^y q(y',t)\,\mathrm{d}y' - \int_0^t
    c(0,t')\,\mathrm{d} t' ,
  \end{split}
\end{equation}
where the soliton speed satisfies $c(y,t) = a(y,t)/3 + q(y,t)^2 =
-\xi_t(x,y,t)$ (cf.~\eqref{eq:KPsoliton} and \eqref{eq:56}).

The class of initial data we consider here corresponds to expansive
conditions, so that we are guaranteed global existence of modulation
solutions.  We will obtain explicit modulation solutions in the form
of simple waves and their interactions that describe the evolution of
the data shown in Fig.~\ref{fig:initial_data}.

Our analysis is supported by numerical simulations of the KPII
equation (\ref{eq:KP}) using a Fourier pseudospectral method adapted
from \cite{kao_numerical_2012} that allows for outgoing line solitons
at the top and bottom of the simulation domain
$[-L_x,L_x]\times[-L_y,L_y]$ through use of a windowing function.  We
maintain the nonlocal constraint $\int_{-L_x}^{L_x}
u_{yy}\,\mathrm{d}x = 0$ to high accuracy by including localised
``image'' initial data whose superposition with the test data of
interest satisfy this constraint.  Simulations were terminated before
the image and test data interacted.  For further details, see Appendix
\ref{sec:numerical-method-1}.

\section{Basic properties of the modulation equations}
\label{sec:expans-init-cond}

In this section, we summarise the classical analysis of the hyperbolic
system (\ref{eq:modulationEqs}).  

The modulation equations admit the amplitude symmetry
\begin{subequations}
  \label{eq:35}
  \begin{equation}
    \label{eq:31}
    a' = a/A, \quad q' = q/\sqrt{A}, \quad y' = \sqrt{A} y, \quad t' =
    t, \quad A > 0 ,
  \end{equation}
  the quasi-rotational symmetry
  \begin{equation}
    \label{eq:34}
    a' = a, \quad q' = q + Q, \quad y' = y - 2Q t, \quad t' = t, \quad
    Q\in \R,
  \end{equation}
  the hydrodynamic scaling symmetry
  \begin{equation}
    \label{eq:32}
    y' = \alpha y, \quad t' = \alpha t, \quad \alpha > 0,
  \end{equation}
  and the reflection symmetry
  \begin{equation}
    \label{eq:33}
    y' = -y, \quad q' = -q ,
  \end{equation}
\end{subequations}
all leaving (\ref{eq:modulationEqs}) unchanged in primed coordinates.
Namely, if $a(y,t)$ and $q(y,t)$ solve \eqref{eq:modulationEqs}, so do
$a'(y',t')$ and $q'(y',t')$.

It was shown in \cite{biondini_integrability_2019} that appropriate
linear combinations of equations (\ref{eq:25}) and (\ref{eq:27})
result in the equivalent pair of equations in characteristic form
\begin{equation}
  \label{eq:30}
  \pm \left [ a_t + (2q \pm \tfrac{2}{3}\sqrt{a})a_y\right ] + 2
  \sqrt{a} \left [ q_t + (2q \pm \tfrac{2}{3}\sqrt{a})q_y\right ] = 0,
\end{equation}
which reveal the characteristic velocities $U = 2q - \frac{2}{3}
\sqrt{a}$ and $V = 2q + \frac{2}{3} \sqrt{a}$.  Integration of
(\ref{eq:30}) along each characteristic direction demonstrates that
\begin{equation}
  \label{eq:1}
  r = q - \sqrt{a}, \quad s = q + \sqrt{a}
\end{equation}
are Riemann invariants for the modulation equations
(\ref{eq:modulationEqs}), which afford the diagonalisation
\begin{subequations}
  \label{eq:2}
  \begin{gather}
    \label{eq:RI_y}
    r_t + U r_y = 0, \quad s_t + V s_y = 0, \\
    \label{eq:67}
    U = \frac{2}{3}(2r + s), \quad V = \frac{2}{3}(r + 2s) ,
  \end{gather}
\end{subequations}
where the characteristic velocities $U$ and $V$ are now written in
terms of the Riemann variables $r$ and $s$.

Simple wave solutions of \eqref{eq:2} correspond to variation in only
one characteristic direction so that one of the Riemann invariants $r$
or $s$ is constant, i.e., either $q(y,t) + \sqrt{a(y,t)}$ or
$q(y,t) - \sqrt{a(y,t)}$ is independent of $y$ and $t$.  Since the
characteristic velocities are ordered $U \le V$, we identify simple
waves with variation along the slow characteristics
$\mathrm{d}x/\mathrm{d}t = U$ as 1-waves ($s = const$) and those with
variation along the fast characteristics $\mathrm{d}x/\mathrm{d}t = V$
as 2-waves ($r = const$).  Across a simple wave, the non-constant
Riemann invariant's characteristic velocity is monotonically
increasing because the strictly hyperbolic system
\eqref{eq:modulationEqs} is genuinely nonlinear so long as $a \ne 0$
\cite{biondini_integrability_2019}.

Expansive initial data $a(y,0)$, $q(y,0)$ corresponds to the condition
that both characteristic velocities $U$ and $V$ evaluated on the
initial data are monotonically increasing functions of $y$.  By virtue
of the fact that $\partial U/\partial r = \partial V/\partial s =
\frac{4}{3} > 0$, expansive initial data corresponds to the case where
both Riemann invariants $r$ and $s$ are non-decreasing functions of
$y$.

Simple waves propagate into constant regions of the $y$-$t$ plane, and
are therefore fundamental building blocks for Riemann problems that
posit step initial data at the origin.  The interaction of simple
waves can most conveniently be investigated by use of the hodograph
transformation \cite{courant_supersonic_1948} in which the role of
dependent and independent variables is swapped.  Namely, we take $t =
t(r,s)$, $y = y(r,s)$, yielding the following set of linear equations
\begin{subequations}
  \label{eq:3}
  \begin{gather}
    \label{eq:28}
    t_{rs} + \frac{2}{s-r}(t_r - t_s) = 0, \\
    \label{eq:29}
    y_s = U t_s , \quad y_r = V t_r, 
  \end{gather}
\end{subequations}
so long as the Jacobian $J = r_t s_y - s_t r_y$ remains nonzero.
Equation (\ref{eq:28}) is an Euler-Poisson-Darboux (EPD) equation that
is equivalent to the radial wave equation in five dimensions, which
admits the general solution
\begin{equation}
  \label{eq:37}
  t(r,s) = A + \left ( \frac{F(r)}{(s-r)^2} \right )_r + \left (
    \frac{G(s)}{(s-r)^2} \right )_s ,
\end{equation}
for an arbitrary constant $A \in \R$ and functions $F(r)$, $G(s)$.  In
order to determine $A$, $F$, and $G$, one must specify suitable
initial and/or boundary conditions.

\section{Partial and truncated line solitons}
\label{sec:truncated-soliton}

By a partial or truncated line soliton, we mean initial data in which
the soliton modulation amplitude is zero on one or two semi-infinite
intervals, respectively.  This scenario where $a = 0$ corresponds to a
vacuum state in which the soliton slope $q$ is undefined.  To resolve
this ambiguity, we will require that the value of $q$ in the
neighbourhood of the point where $a$ becomes positive corresponds to a
simple wave in which one of $r$ or $s$ is constant
\cite{smoller_shock_1994}.  Then, the propagation speed of the vacuum
front will necessarily be $\lim_{a\to 0} U = \lim_{a\to 0} V = 2q$.

\subsection{Partial line solitons}
\label{sec:partial-half-line}

This problem was previously posed and solved in
\cite{neu_singular_2015} as a model for two-dimensional soliton
diffraction.  Here, we show that the resulting simple wave solution
forms an important building block for other, more complex, truncated
and partial soliton interactions.

Without loss of generality (recall the symmetries (\ref{eq:35})), we consider the partial (half) line soliton initial data
\begin{equation}
  \label{eq:36}
  a(y,0) =
  \begin{cases}
    0 & y > 0 \\ 1 & y \le 0
  \end{cases}, \quad q(y,0) = 0, \quad y < 0,
\end{equation}
for the modulation equations (\ref{eq:modulationEqs}).  This Riemann
problem is solved by a 1-wave corresponding to a centred,
self-similar rarefaction wave (1-RW) in which $s = q+\sqrt{a} \equiv
1$ and $U = y/t$ \cite{neu_singular_2015}
\begin{equation}
  \label{eq:38}
    \sqrt{a_{\rm p}(y,t)} =
    \begin{cases}
      0 & \quad 2t < y \\ \frac{3}{4} \left ( 1- \frac{y}{2t} \right ) &
      -\frac{2}{3} t < y < 2t \\ 1 & \qquad\quad\, y < -\frac{2}{3} t
    \end{cases}, \quad q_{\rm p}(y,t) = 1 - \sqrt{a_{\rm p}(y,t)} .
\end{equation}
This solution describes the progressive disintegration of the partial
soliton.
\begin{figure}
  \centering
  \includegraphics{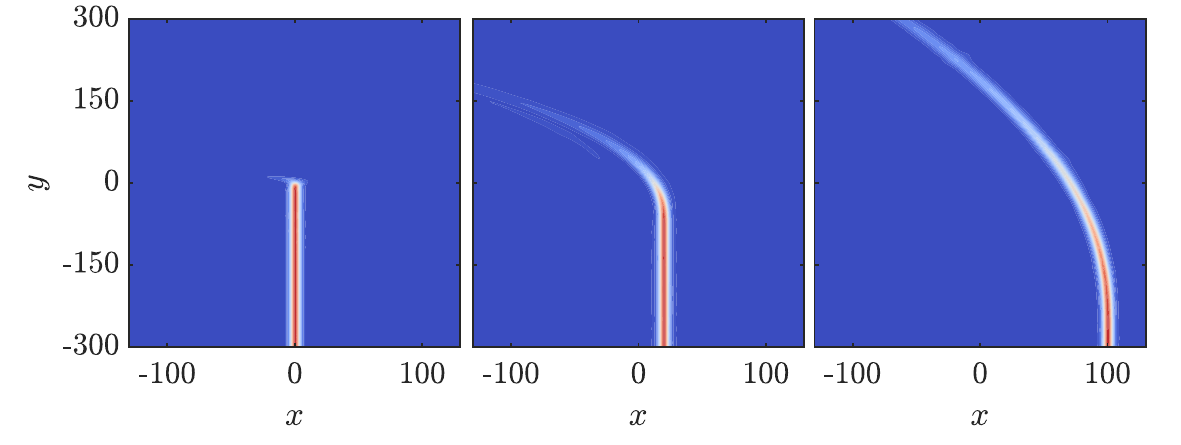}
  \includegraphics[scale=.25]{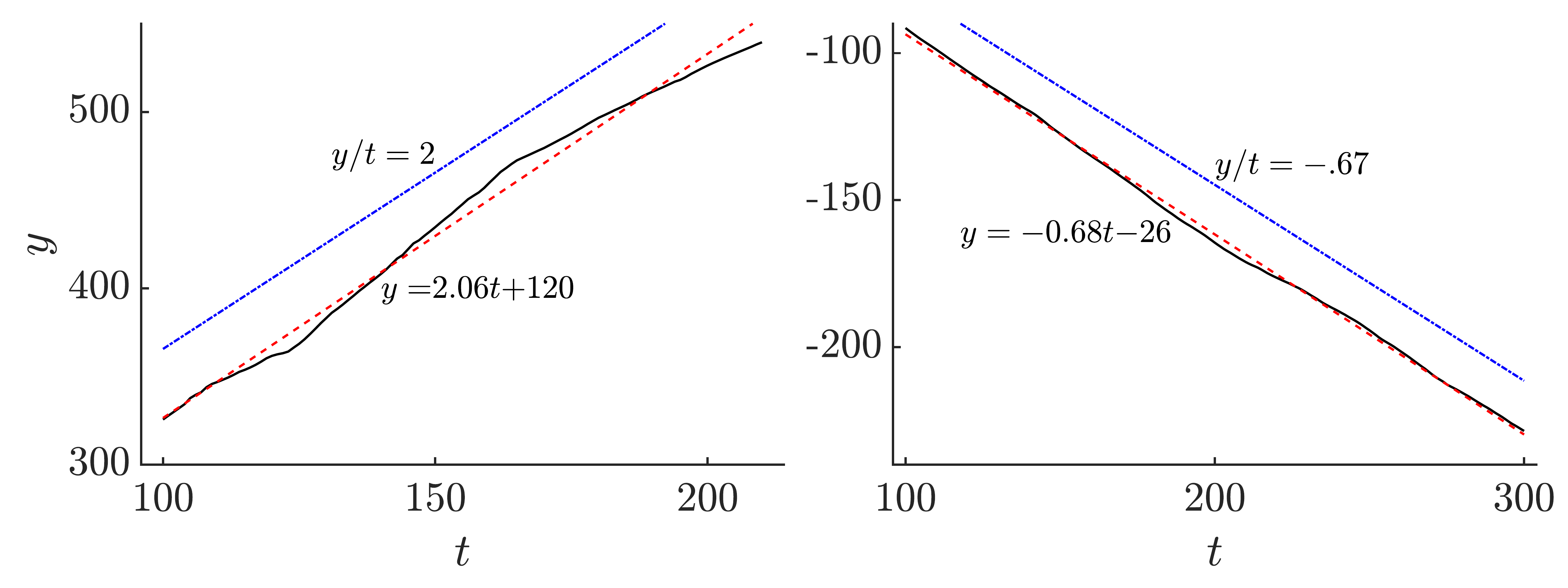}
  \caption{Top: numerical simulation of partial soliton evolution for
    $t \in (0,60,300)$.  Bottom: comparison of characteristic speeds
    of the upper (left panel) and lower (right panel) edges of the
    partial soliton rarefaction wave. The plot displays the predicted
    front speeds as reference lines (dash-dotted, blue) with slopes
    from the solution \eqref{eq:38}, the numerically extracted front
    positions (solid, black), and a least squares linear fit (dashed,
    red) whose slope determines the measured speeds.}
  \label{fig:partial_soliton}
\end{figure}
These dynamics are shown in Figure \ref{fig:partial_soliton} top,
where we depict the evolution of the partial soliton, as computed via
numerical integration of the KP equation (see
App.~\ref{sec:numerical-method-1}). In Fig.~\ref{fig:partial_soliton}
bottom, we compare the analytical predictions with numerical evolution
by identifying the positions of the vacuum front and vertical soliton
front as the first points where $a \to 0.03$ and $a \to 0.97$,
respectively.  These fronts are well approximated by straight lines
with slopes given by the predicted characteristic speeds from the
solution in Eq.~\eqref{eq:38}.  Below, we will use the solution in
Eq.~\eqref{eq:38} to analyse the evolution resulting from more complex
initial conditions.

\subsection{Truncated line solitons}
\label{sec:trunc-line-solit}

We now consider the KPII equation with initial conditions for a
truncated line soliton (recall Fig.~\ref{fig:initial_data} left) of
length $\ell > 0$ by imposing the modulation initial conditions
\begin{equation}
    \label{eq:TruncatedIC}
    a(y,0)=
    \begin{cases} 
        1 & |y| \le \ell/2 \\
        0 & |y| > \ell/2
    \end{cases}, \quad
    q(y,0)= 0, \quad |y| < \ell/2 .
\end{equation}
By the reflection symmetry \eqref{eq:33}, $q$ and $a$ are respectively
odd and even functions of $y$ for each $t \ge 0$.  As in the case of
the partial soliton, the truncated soliton slope in the vacuum region
where $a = 0$ is determined by a simple wave condition.  Namely, for
short times, a non-centred 1-RW is generated from the upper truncation
point at $y = \ell/2$.  Similarly, a 2-RW is generated from the lower
truncation point $y = -\ell/2$.  By use of the reflection symmetry
(\ref{eq:33}), the initial evolution of these two simple waves can be
represented as even or odd extensions of a shifted partial soliton
\eqref{eq:38}
\begin{equation}
  \label{eq:40}
  a(y,t) = a_{\rm p}(|y|-\ell/2,t), \quad q(y,t) = \mathrm{sgn}(y)q_{\rm
    p}(|y|-\ell/2,t),
\end{equation}
for $y \in \R$.  However, this solution only holds prior to the
interaction of the simple waves, which limits its validity to $0 \le t \le
\frac{3}{4} \ell$.  A characteristic diagram showing the 1-RW and 2-RW
solutions emanating from $Y = y/\ell = \pm 1/2$ is shown in
Fig.~\ref{fig:truncated_characteristics}.

At $t = \frac{3}{4}\ell$, the two simple waves intersect at $y = 0$.
In order to understand what happens for $T = t/\ell > \frac{3}{4}$, we
utilise the hodograph transformation and the corresponding equations
(\ref{eq:3}).  The boundary conditions for the EPD equation
(\ref{eq:28}) can be obtained by recognising that, at the boundaries
of the simple wave interaction region, either $r$ or $s$ is constant.
When $s = 1$, as in the 1-RW propagating down from $y = \ell/2$, we
differentiate the simple wave equation
\begin{equation}
  \label{eq:15}
  \frac{y-\ell/2}{t} = U = \frac{2}{3}(2r + 1)
\end{equation}
with respect to $r$ to obtain the relation
\begin{equation}
  \label{eq:16}
  y_r = \frac{2}{3}t_r(2r+1) + \frac{4}{3} t .
\end{equation}
Using this expression to eliminate $y_r$ from Eq.~(\ref{eq:29}), we
obtain
\begin{subequations}
  \label{eq:41}
  \begin{equation}
    \label{eq:17}
    t_r - \frac{2}{1-r} t = 0 , \quad s = 1 , \quad r \in [-1,1] .
  \end{equation}
  Likewise, for the other 2-RW propagating up from $y = -\ell/2$, we
  obtain
  \begin{equation}
    \label{eq:18}
    t_s + \frac{2}{1+s} t = 0, \quad r = -1, \quad s \in [-1,1] .
  \end{equation}
\end{subequations}
Integrating \eqref{eq:41} from the initial time of simple wave
interaction $t(-1,1) = 3\ell/4$, we obtain the boundary conditions
\begin{subequations}
  \label{eq:42}
  \begin{align}
    \label{eq:19}
    t(r,1) &= \frac{3\ell}{(1-r)^2}, \quad r \in [-1,1), \\
    \label{eq:44}
    t(-1,s) &= \frac{3\ell}{(1+s)^2}, \quad s \in (-1,1] ,
  \end{align}
\end{subequations}
for Eq.~(\ref{eq:28}).  Applying these boundary conditions to the
general solution (\ref{eq:37}) yields $A = 0$, $F(r) =
\frac{3}{4}\ell(2-(1-r)^2)$, $G(s) = -\frac{3}{4}\ell(2-(1+s)^2)$ and
the solution
\begin{subequations}
  \label{eq:68}
  \begin{equation}
    \label{eq:13}
    t(r,s) = 3 \ell \frac{(1-rs)}{(s-r)^3}, \quad (r,s) \in [-1,1)\times
    (-1,1] .
  \end{equation}

  \begin{figure}
    \centering
    \includegraphics[scale=0.25]{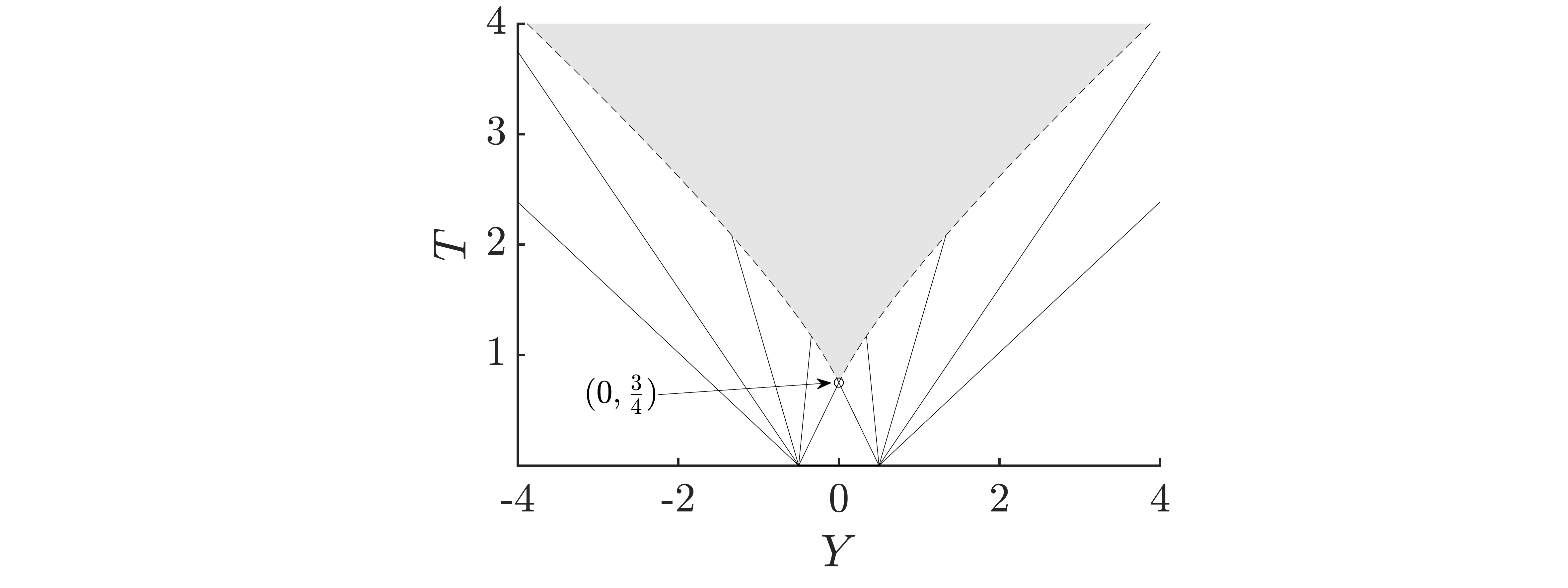}
    \caption{Simple waves and their interaction in the characteristic
      $Y$-$T$ ($\frac{y}{\ell}$-$\frac{t}{\ell}$) plane for the
      truncated soliton initial data \eqref{eq:TruncatedIC}.  The
      solid lines are simple wave characteristics.  The shaded region
      corresponds to interacting simple waves bounded by the dashed
      curves from Eq.~\eqref{eq:7}.}
    \label{fig:truncated_characteristics}
  \end{figure}
  We now solve for $y(r,s)$, by integrating both of Eq.~(\ref{eq:29})
  \begin{equation}
    \label{eq:20}
    y(r,s) = \frac{\ell(r+s)(r^2 + 4 r s + s^2 - 6)}{2(r-s)^3}, \quad
    (r,s) \in [-1,1]^2 ,
  \end{equation}
\end{subequations}
where we used $y(-1,1) = 0$ at the initiation of simple wave
interaction.  The expressions \eqref{eq:68} implicitly determine the
simple wave interaction for $t \ge \frac{3}{4}\ell$.

We observe that the quantities
\begin{equation}
  \label{eq:6}
  Y = \frac{y}{\ell}, \quad T = \frac{t}{\ell}
\end{equation}
are independent of the truncated soliton length $\ell$, a
manifestation of the hydrodynamic symmetry (\ref{eq:32}).  We will
henceforth report results in the scaled variables $Y$ and $T$.  

The boundary of the simple wave interaction region is determined by
evaluating the hodograph solution \eqref{eq:68} at $s = 1$ for $Y > 0$
and using reflection symmetry
\begin{equation}
  \label{eq:7}
  |Y| = \frac{1}{2} + 2T - \frac{4}{3} \sqrt{3 T}, \quad T
  \ge \frac{3}{4} .
\end{equation}
These are the dashed curves in
Fig.~\ref{fig:truncated_characteristics}.

As shown in Fig.~\ref{fig:truncated_characteristics} for short times,
two non-centred simple waves described by (\ref{eq:38}),
(\ref{eq:40}) emanate from the soliton truncation points at $Y = \pm
\frac{1}{2}$.  For long times, the interaction boundary (\ref{eq:7})
approaches $|Y| \sim 2 T$ with the same slope as the outermost edges
of the simple waves $|Y| = 2T + 1$.  Note, however that the two
characteristic curves never cross.

Returning to the physical variables $a$ and $q$ using \eqref{eq:1},
the simple wave interaction is described by the hodograph solution
\eqref{eq:68}, which yields the expressions
\begin{subequations}
  \label{eq:24}
  \begin{align}
    \label{eq:72}
    Y &= \frac{q}{4 a^{3/2}}(3+a-3q^2), \\
    \label{eq:73}
    T &= \frac{3}{8a^{3/2}}(1+a-q^2) .
  \end{align}
\end{subequations}
Since $q(0,T) \equiv 0$ from \eqref{eq:72}, we can obtain the explicit
decay of the soliton amplitude at the origin from \eqref{eq:73}
\begin{subequations}
  \label{eq:69}
  \begin{align}
    \label{eq:12}
    \sqrt{a(0,T)} &= \frac{1}{8T}\left ( 2 f(T)^{1/3} + 1 +
      \frac{1}{2} f(T)^{-1/3}
    \right ), \quad \frac{3}{4} \le T , \\
    \label{eq:70}
    f(T) &= \frac{1}{8} + 12 T^2 + T \sqrt{3 + 144 T^2} .
  \end{align}
\end{subequations}
Using \eqref{eq:24}, one could obtain explicit expressions for
$a = a(Y,T)$ and $q = q(Y,T)$ for general $Y$ and $T$.  However, we
can draw several important conclusions from the asymptotics of the
implicit solution (\ref{eq:24}).  For $T \gg 1$ and $|Y| \ll T^{2/3}$,
the amplitude is approximately independent of $Y$ and the slope is
approximately linear in $Y$
\begin{equation}
  \label{eq:9}
  \begin{split}
    a(Y,T) &\sim \frac{1}{4} \left ( \frac{3}{T} \right )^{2/3} +
    \frac{3^{1/3}}{8} \left ( \frac{1}{T} \right )^{4/3}, \\ 
    q(Y,T) &\sim \frac{Y}{2 T} + \frac{Y}{4\cdot 3^{1/3}} \left (
      \frac{1}{T} \right )^{5/3} .
  \end{split}
\end{equation}
The one-term expansion for $a$ and $q$ in \eqref{eq:9} is a
self-similar solution of the modulation equations \eqref{eq:modulationEqs} \cite{lee_upstreamadvancing_1990}.

Using the above modulation solution to reconstruct the approximate
soliton (\ref{eq:KPsoliton}) yields interesting predictions for the
initial data
\begin{equation}
  \label{eq:8}
  u(x,y,0) =
  \begin{cases}
    \mathrm{sech}^2 \left ( \sqrt{\frac{1}{12}}x \right ) & |y| \le
    \frac{\ell}{2} \\
    0 & |y| > \frac{\ell}{2} 
  \end{cases}
\end{equation}
to the KPII equation (\ref{eq:KP}).  The soliton phase
$\xi = \xi(x,y,t)$ in Eq.~(\ref{eq:KPsoliton}) can be approximated for
large $t$ using (\ref{eq:9}) as
\begin{equation}
  \label{eq:23}
  \xi(x,y,t) \sim x+\frac{y^2}{4t}-\left( \frac{3\ell}{8}
  \right)^{2/3} t^{1/3} 
\end{equation}
Because the soliton maximum occurs where $\xi=0$, we conclude that,
for long times, the truncated line soliton shape approaches a moving
parabola opening in the negative $x$ direction with increasing focal
length $t$
\begin{equation}
    \label{eq:s3}
    x+\frac{y^2}{4t}=c(t)t, \quad
    c(t)=\left(\frac{3\ell}{8t}\right)^{2/3} .
\end{equation}
While the parabolic shape is independent of the initial truncated
soliton length $\ell$, the wave speed is proportional to $\ell^{2/3}$.
Concurrently, the soliton amplitude decays according to
Eq.~(\ref{eq:9}).

The shape and amplitude of the modulated line soliton within the
simple wave interaction region has an interesting connection to the
cylindrical KdV (cKdV) equation. As noted in
\cite{ablowitz_dispersive_2016}, introducing the change of variables
\begin{equation}
  \label{eq:10}
  u(x,y,t) = f(\eta,t), \quad \eta = x + \frac{y^2}{4t}
\end{equation}
results in the exact reduction of the KPII equation (\ref{eq:KP}) to
the cKdV equation
\begin{equation}
    \label{eq:cKdV}
    f_t+ff_\eta+f_{\eta\eta\eta}+\frac{1}{2t} f =0.
\end{equation}
This equation admits slowly decaying soliton solutions
\cite{nakamura_soliton_1981}.  Approximate soliton solutions for $t
\gg 1$ take the form of slowly varying KdV solitons
\cite{ko_cylindrical_1979}
\begin{equation}
  \label{eq:11}
  f(\eta,t) \sim A(t)\, \mathrm{sech}^2 \left ( \sqrt{\frac{A(t)}{12}}
    (\eta - z(t))  \right ), \quad \dot{z}(t) = \frac{A(t)}{3} .
\end{equation}
In order to determine the slowly varying amplitude $A(t)$, we appeal
to the conserved momentum
\begin{equation}
  \label{eq:14}
  P = \int_{\R} t f(\eta,t)^2 \, \mathrm{d}\eta
\end{equation}
for any square integrable solution of cKdV (\ref{eq:cKdV}).  In
shallow water waves, the quantity $P$ is identified with the momentum
because $\sqrt{t} f$ is proportional to both the deviations of water
height and vertically averaged horizontal velocity from their
equilibrium values for cylindrical waves
\cite{johnson_modern_1997}. Inserting the slowly varying soliton
ansatz (\ref{eq:11}) into (\ref{eq:14}), we obtain
\begin{equation}
  P = t\frac{8A(t)^{3/2}}{\sqrt{3}} .
\end{equation}
Thus, given some momentum $P_0$, the slowly varying amplitude is
\begin{equation}
  \label{eq:21}
  A(t) = \left ( \frac{3^{1/3} P_0}{8 t} \right )^{2/3} .
\end{equation}
If we choose the initial momentum to be $P_0 = 9 \ell$, then this
amplitude equation matches the leading order truncated soliton
amplitude $a(Y,T)$ (\ref{eq:9}) for large $T$.  Moreover, the
approximate cKdV soliton (\ref{eq:11}) admits the phase
\begin{equation}
  \label{eq:22}
  \eta - z(t) = x + \frac{y^2}{4 t} - \left ( \frac{3 \ell}{8} \right
  )^{2/3} t^{1/3},
\end{equation}
which also matches the truncated soliton phase in Eq.~(\ref{eq:23}),
i.e., the leading order soliton slope $q(Y,T)$ in (\ref{eq:9}).

\begin{figure}
  \centering
  \includegraphics{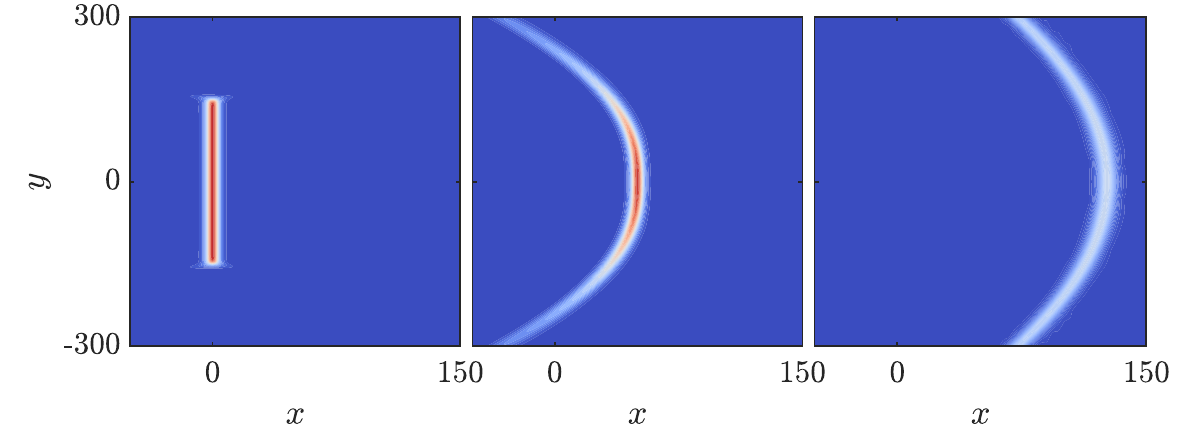}
  \includegraphics[scale=0.25]{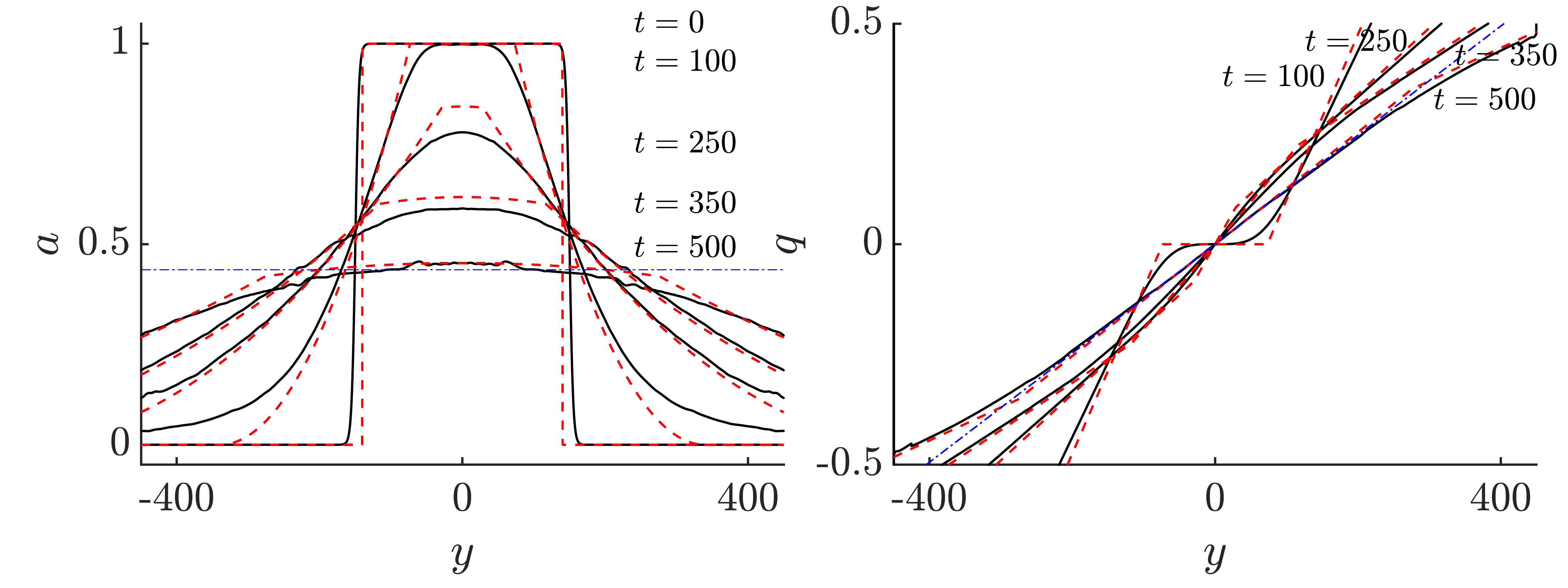}
  \caption{Top: numerical evolution of truncated soliton initial data
    \eqref{eq:61} according to the KPII equation for $t \in
    (0,150,500)$ and $\ell = 300$.  Bottom: modulated soliton
    amplitude $a$ and slope $q$ at noted times extracted from the
    numerical simulation (solid curves) and the modulation solution
    (\ref{eq:40}), (\ref{eq:24}) (dashed curves) with the slightly
    different, fitted initial soliton length $\ell = 280$.  The
    dash-dotted blue lines correspond to the long-time asymptotic
    predictions (\ref{eq:9}) evaluated at $t = 500$.}
  \label{fig:truncated_numerics}
\end{figure}
Figure \ref{fig:truncated_numerics} top depicts a numerical simulation
of the truncated soliton initial data (\ref{eq:8}) with $\ell = 300$
that has been smoothed so as to minimise Gibbs type oscillations
\cite{biondini_gibbs_2017}.  Curved waves emanate from the truncation
edges as the central portion propagates forward.  When the central
prominence decays, the entire wave forms a curved shape with decaying
amplitude and curvature as time increases.  These qualitative features
are reflected in the obtained modulation solution for the soliton
amplitude $a(y,t)$ and slope $q(y,t)$ in Equations \eqref{eq:40},
\eqref{eq:24}.  In order to quantitatively compare the simulation to
modulation theory predictions, we extract the modulated soliton
amplitude and slope from the simulation via
\begin{equation}
  \label{eq:60}
  a(y,t) = \max_{x \in \R} u(x,y,t), \quad q(y,t) = -\left (
    \argmax_{x \in \R} u(x,y,t) \right )_y .
\end{equation}
For the numerical computation of $q$, we smooth $\argmax u$ prior to
differentiation.  Figure \ref{fig:truncated_numerics} bottom displays
the numerical (solid) and modulation (dashed) solutions.  In order to
quantitatively track the numerical simulation, we used the slightly
smaller truncation width $\ell = 280$ for the modulation solution in
order to account for the smoothing of the initial data as given in
(\ref{eq:61}).  Both the soliton amplitude and slope closely match the
full PDE evolution described by Eq.~\eqref{eq:KP}, demonstrating that
our modulation analysis captures both the qualitative and quantitative
features of the solution.  The long-time ($T \gg 1$) asymptotic
predictions in Eq.~(\ref{eq:9}) (dash-dotted) for a parabolic,
decaying cKdV soliton also compare favourably with the numerical and
modulation solutions for $|y| \lessapprox \ell$ despite the modest
scaled time $T = t/\ell \approx 1.5$.

In summary, the truncated soliton initial data (\ref{eq:8}) evolves
into a curved soliton with algebraically decaying amplitude that
approaches a cKdV soliton with a parabolic profile and linearly
increasing focal length.


\section{Bent-stem and bent line solitons}
\label{sec:kink-kink-stem}

The modulation solution for the truncated soliton consisting of two
counterpropagating and then interacting simple waves motivates a
broader class of initial conditions where we relax the assumption of
zero soliton amplitude for $|y| > \ell/2$.  In this section, we
explore this scenario with three distinct configurations: a special
partially bent soliton, two bent solitons joined via a larger
amplitude stem with nonzero $\ell$, and finally a bent soliton with
the same amplitude throughout in which $\ell \to 0$---the regular and
Mach expansion problem.  First we consider the partially bent soliton
as a natural extension of the partial line soliton described in
Sec.~\ref{sec:partial-half-line}.

\subsection{Partially bent solitons}
\label{sec:partial-bent-stem}

\begin{figure}
  \centering
  \includegraphics[]{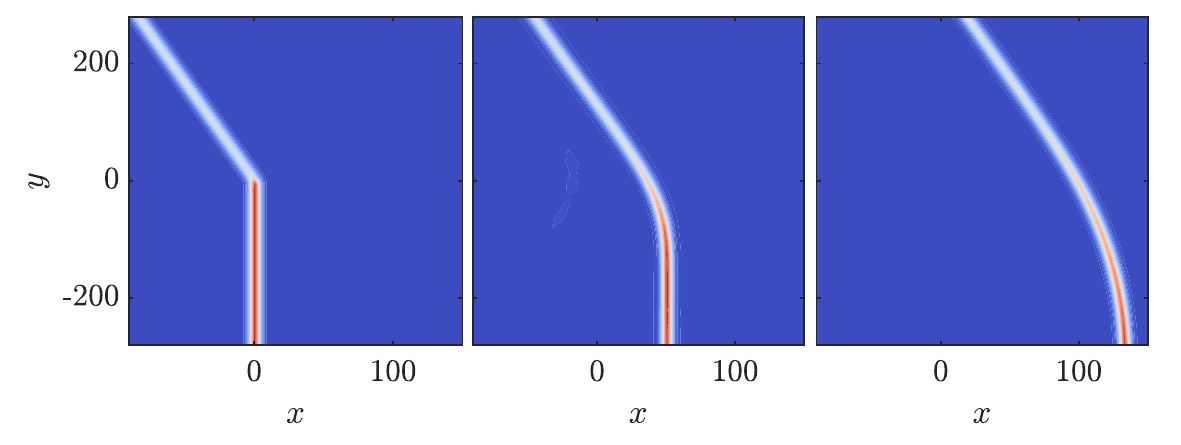}
  \includegraphics[scale=.25]{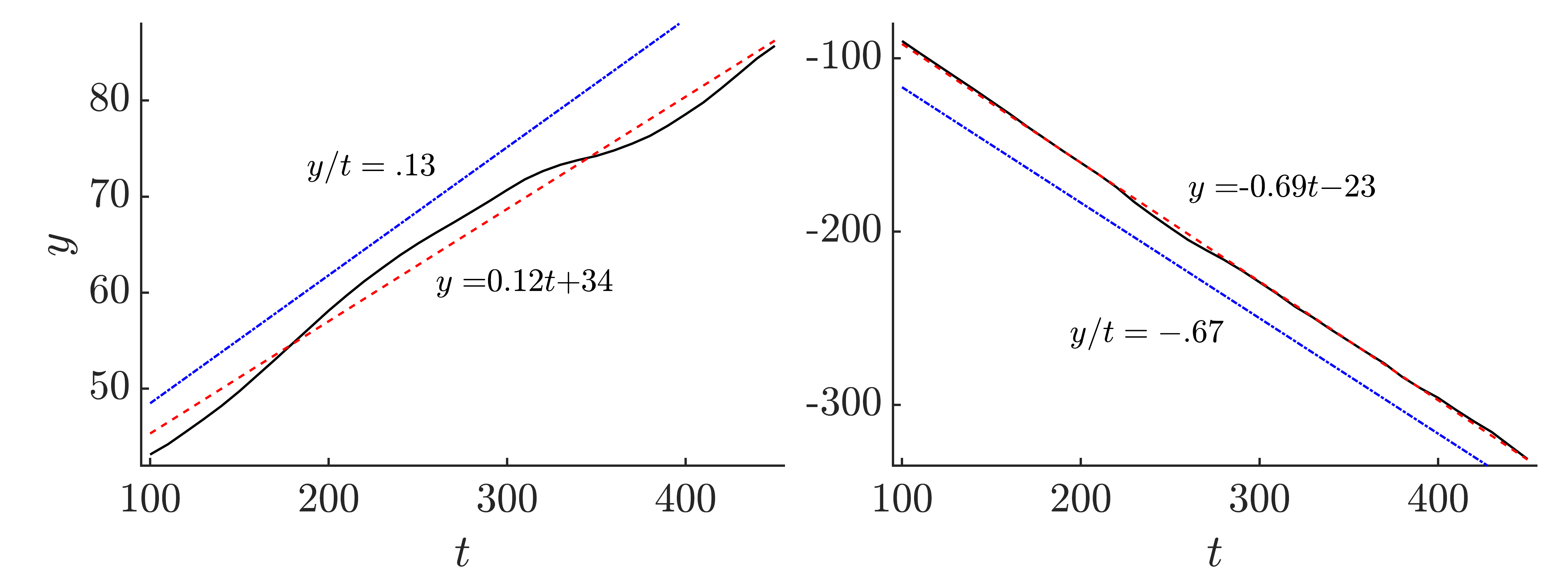}
  \caption{Top: numerical evolution of the partially bent soliton
    according to the KPII equation for $t \in (0,150,400)$ and
    $\sqrt{a_0} = 0.7$, $q_0 = 0.3$. Bottom: comparison of the characteristic
    speeds of the upper (left panel) and lower (right panel) edges of
    the partially bent soliton rarefaction wave. The plot displays the
    predicted front speeds from the modulation solution \eqref{eq:62}
    as reference lines (dash-dotted, blue), the numerically extracted
    front positions from the numerical simulation (solid, black), and
    a least squares linear fit (dashed, red) whose slope determines
    the measured speeds.}
  \label{fig:partial_bent}
\end{figure}

We first consider a single bend at $y = 0$ where
\begin{equation}
  \label{eq:71}
  a(y,0)=
  \begin{cases} 
    1 & y \le 0 \\
    a_0 & y > 0 
  \end{cases}, \quad
  q(y,0)=
  \begin{cases} 
    0 & y \le 0 \\
    q_0 & y > 0
  \end{cases}, 
\end{equation}
For generic choices of $0 < a_0 < 1$ and $0 < q_0 < 1$, the initial
data \eqref{eq:71} give rise to two separated simple wave solutions of
the modulation equations \eqref{eq:modulationEqs}: a fast 2-RW and a
slow 1-RW separated by a constant region.  As a natural extension of
the partial line soliton solution \eqref{eq:38}, we restrict the data
\eqref{eq:71} so that only a single simple wave---the 1-RW---is
generated, i.e., $s = q + \sqrt{a}$ is constant:
\begin{equation}
  \label{eq:26}
  q_0 + \sqrt{a_0} = 1, \quad 0 < a_0 < 1, \quad 0 < q_0 < 1 .
\end{equation}
We call the corresponding initial data \eqref{eq:71} subject to
\eqref{eq:26} a partially bent soliton, which will be useful to
describe the bent-stem soliton initial data in the next subsection.
The constancy of $s$ ($q(y,t) + \sqrt{a(y,t)} = 1$) and $U(a,q) = y/t$
result in the 1-RW solution
\begin{equation}
  \label{eq:62}
  \begin{split}
    \sqrt{a_{\rm pb}(y,t)} &=
    \begin{cases}
      a_0 & ~ U_0 t < y \\ 
      \frac{3}{4} \left ( 1- \frac{y}{2t} \right ) &
      -\frac{2}{3} t < y < U_0 t \\ 
      1 & \qquad\quad\, y < -\frac{2}{3} t
    \end{cases}, \\
    q_{\rm pb}(y,t) &= 1 - \sqrt{a_{\rm pb}(y,t)}, \quad U_0 =
    2 - \frac{8}{3}\sqrt{a_0} .
  \end{split}
\end{equation}
This solution is the same as that of the partial soliton (\ref{eq:38})
for $y < U_0 t$ and limits to the partial soliton modulation as the
angled soliton amplitude vanishes $a_0 \to 0$.  The positive
amplitude, outgoing soliton gives rise to the slower characteristic
velocity $U_0 < 2$.  The evolution of a partially bent soliton
\eqref{eq:71} with $\sqrt{a_0} = 0.7$, $q_0 = 0.3$ according to numerical
integration of the KP equation \eqref{eq:KP} is shown in Figure
\ref{fig:partial_bent} top.  The 1-RW modulation solution's edge
characteristic velocities $U_0$ and $-\frac{2}{3}$ in
Eq.~\eqref{eq:62} are favourably compared with the numerical
simulation in Fig.~\ref{fig:partial_bent} bottom by identifying the
front positions where $a \to 0.52$ and $a \to 0.97$, respectively.

The solution \eqref{eq:62} provides a building block to analyse the
more complicated configuration of a bent-stem soliton.

\subsection{Bent-stem solitons}
\label{sec:kink-stem-soliton}

We now consider the initial condition
\begin{equation}
    \label{eq:KinkIC}
    a(y,0)=
    \begin{cases} 
        1 & |y| \le \ell/2 \\
        a_0 & |y| > \ell/2 
    \end{cases}, \quad
    q(y,0)=
    \begin{cases} 
        0 & |y| \le \ell/2 \\
        \mathrm{sgn}(y) q_0 & |y| >\ell/2
    \end{cases}, 
\end{equation}
which is the modulation initial condition for the KPII data depicted
in Fig.~\ref{fig:initial_data} middle.  This configuration describes
an initial truncated soliton of length $\ell$ that is extended with
outgoing line solitons of amplitude $a_0 < 1$ and nonzero, symmetric
slopes $\pm q_0$.  The case $a_0 = 0$ and $q_0 = 1$ corresponds to the
truncated soliton \eqref{eq:TruncatedIC}.  As similarly noted for the
partially bent soliton, generic choices of $0< a_0 < 1$ and
$0 < q_0 < 1$ will give rise to four separated non-centred simple
waves, two at each bend $y = \pm \ell/2$.  The fastest and slowest
waves, however, will not interact with the other waves, propagating
far away from the initial stem region.  These non-interacting,
propagating waves are of less interest so we restrict the initial data
such that a single simple wave is generated at each bend, as in the
partially bent soliton case.  Consequently, we assume the same simple
wave constraint in Eq.~\eqref{eq:26} corresponding to a non-centred
1-RW emanating from $y = \ell/2$ and a non-centred 2-RW emanating from
$y = -\ell/2$.  We call the corresponding initial data
(\ref{eq:KinkIC}) a bent-stem soliton.

This initial value problem is nearly identical to the truncated
soliton problem.  In fact, their solutions are essentially the same
apart from one subtle yet crucial difference: the velocities of the
outermost edges of the counterpropagating simple waves are
different. These differing velocities lead to different interaction
features.

We now use the partially bent soliton simple wave \eqref{eq:62} to
construct the counterpropagating simple waves for the bent-stem
soliton initial data (\ref{eq:KinkIC})
\begin{equation}
  \label{eq:46}
  a(y,t) = a_{\rm pb}(|y|-\ell/2,t), \quad q(y,t) =
  \mathrm{sgn}(y)q_{\rm pb}(|y|-\ell/2,t),
\end{equation}
for $y \in \R$ prior to simple wave interaction $0 \le t \le
\frac{3}{4}\ell$.  

\begin{figure}
  \centering
  \includegraphics[scale=.25]{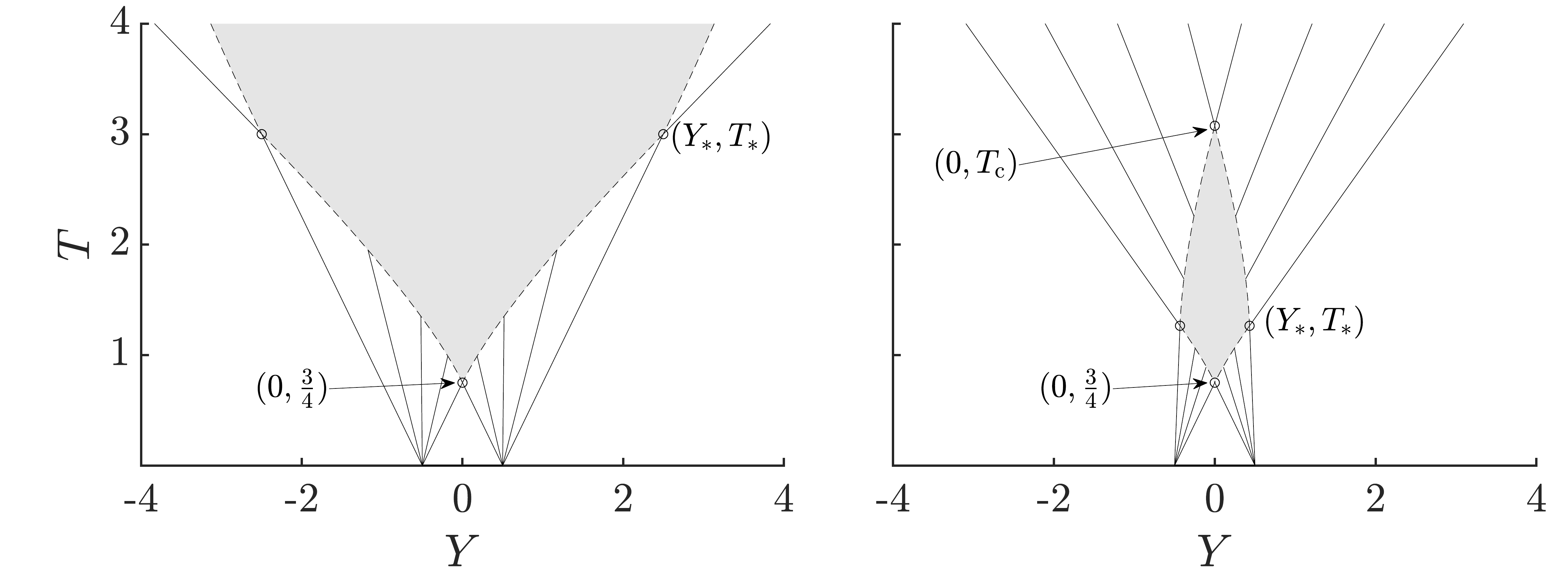}
  \caption{Characteristic plots of interacting simple waves for the
    bent-stem soliton initial data \eqref{eq:KinkIC}.  Left: $0 < a_0
    \le \frac{1}{4}$, resulting in an infinite region of interaction
    for two simple waves in $(y,t)$-plane.  Right: $\frac{1}{4} < a_0
    < 1$, resulting in a bounded interaction region.  See main text
    for description.}
  \label{fig:InteractionRegion}
\end{figure}

Compared to the truncated soliton, the Riemann invariants for the
bent-stem soliton simple waves take values on the smaller square
\begin{equation}
  \label{eq:47}
  (r,s) \in [-1,r_0] \times [-r_0,1] ~ \mathrm{where} ~ r_0 =
  1-2\sqrt{a_0} < 1 .
\end{equation}
Consequently, the hodograph solution for the simple wave interaction
region is the same as for the truncated soliton, namely
Eq.~(\ref{eq:68}).  However, the solution must be considered on the
restricted domain (\ref{eq:47}).  Two space-time characteristic
diagrams of the modulation solution for different values of $a_0$ are
shown in Fig.~\ref{fig:InteractionRegion}.  The interaction region is
shaded grey. The bottom point of the interaction region corresponds to
the initiation of simple wave interaction when $(Y,T) =
(0,\frac{3}{4})$.  Note that the characteristic for the uppermost edge
of the incoming 1-RW, $Y = \frac{1}{2} + U_0 T$, eventually intersects
the edge of the interaction region (\ref{eq:7}) at $(Y_*,T_*)$.
Similarly, the reflected characteristic emanating from $Y =
-\frac{1}{2}$ intersects the interaction region at $(-Y_*,T_*)$.
These intersection points are given by
\begin{equation}
  \label{eq:48}
  Y_* = \frac{1}{2} + \frac{3- 4\sqrt{a_0}}{2a_0} , \quad T_* =
  \frac{3}{4 a_0} 
\end{equation}
and are shown in the characteristic diagrams of
Fig.~\ref{fig:InteractionRegion} for two different choices of
$a_0$. When $a_0 \to 0$, $T_* \to \infty$ and we recover the result
for the truncated soliton in which the colliding simple waves do not
completely intersect one another.  For the bent-stem soliton in which
$0 < a_0 < 1$, the existence of the intersection points $(\pm
Y_*,T_*)$ occurs because the characteristic velocity $U_0$ is slower
than the corresponding characteristic velocity of the truncated
soliton $U_0 < 2$.  This subtle velocity difference leads to a
significant change in the dynamics as we now explain.

For $T > T_*$, the 2-RW that propagated from the lower bend at $y =
-\ell/2$ emerges from the interaction region as a simple wave with
constant $r = r_0$ and expands along the upper, outgoing soliton.  The
uppermost, leading edge portion of the simple wave is the
straight-line characteristic
\begin{equation}
  \label{eq:43}
  Y = V_0 (T - T_*) + Y_*, \quad V_0 = V(r_0,1) =
  \frac{2}{3}(2\sqrt{a_0}+1) .
\end{equation}
The boundary of the interaction region emanating from $(Y_*,T_*)$ now
becomes the parametric curve
\begin{equation}
  \label{eq:49}
  Y = Y(r_0,s), \quad T = T(r_0,s) , \quad s \in [-r_0,1] ,
\end{equation}
where $Y_* = Y(r_0,1)$, $T_* = T(r_0,1)$ and the curve is traversed as
$s$ is decreased from 1.  A new Cauchy problem for the modulation
equations \eqref{eq:modulationEqs} must be solved with data prescribed
along the parametric curve \eqref{eq:49}.  Because the region into
which this Cauchy problem propagates is constant, $(a,q) = (a_0,q_0)$,
it is a simple wave, a 2-wave with $r = r_0$.  The solution is
determined by identifying the characteristics emanating from the
boundary curve \eqref{eq:49}.  Given any $s \in (\max(-r_0,0),1)$
along the boundary curve \eqref{eq:49}, the corresponding
characteristic along which $s$ is constant is the straight line
\begin{equation}
  \label{eq:45}
  Y = \frac{2}{3}(r_0 + 2s)(T-T(r_0,s)) + Y(r_0,s) .
\end{equation}
Example characteristics are shown in Fig.~\ref{fig:InteractionRegion}
right.

A bifurcation occurs in the shape of the interaction region depending on the initial outgoing soliton amplitude $a_0$. For sufficiently large $a_0$, the
interaction boundary (\ref{eq:49}) terminates when $Y = 0$, which from
the hodograph solution (\ref{eq:20}) occurs when $s = -r_0$.
According to the parametric curve (\ref{eq:49}), it would appear that
$s = -r_0$ can occur for any $-1 \le r_0 < 1$.  However, the hodograph
solution for $T$ in Eq.~(\ref{eq:13}) shows that $T \to \infty$ as $s
\to r_0$.  As $s$ is decreased from $1$ in the parametric curve
(\ref{eq:49}), $s$ attains the value $r_0$ before it reaches $-r_0$ if
and only if $r_0 > 0$.  Consequently, the critical value $r_0 = 0$
determines the bifurcation from an unbounded (when $0 \le r_0 \le 1$)
to a bounded (when $-1 < r_0 < 0$) simple wave interaction region.  We
now consider each case in turn.

The characteristic diagram for an unbounded interaction case where
$r_0 < 0$ (equivalent to either condition $0 < a_0 \le 1/4$ or $1/2
\le q_0 < 1$) is shown in Fig.~\ref{fig:InteractionRegion} left. Aside
from the intersecting characteristics at $(\pm Y_*,T_*)$ and the
concomitant simple wave \eqref{eq:45} that emerges from the
interaction region, the bent-stem soliton diagram is similar to the
truncated soliton solution in which $a_0 = 0$
(cf.~Fig.~\ref{fig:truncated_characteristics}).  In fact, the
long-time asymptotic behaviour of the solution is identical to the
truncated line soliton \eqref{eq:9} when $|Y| \le T^{2/3}$ in which
the stem forms a decaying soliton that approaches the parabolic-shaped
cKdV soliton \eqref{eq:11}.

In contrast, when $r_0 < 0$ (equivalent to either $1/4 < a_0 < 1$ or
$0 < q_0 < 1/2$), corresponding to the case of a bounded simple wave
interaction region, the characteristic diagram is significantly
different as in Fig.~\ref{fig:InteractionRegion} right.  By symmetry,
the interaction boundary must close at $Y_{\rm c} = 0$.  Since
$s = -r_0$ determines the terminus of interaction, the corresponding
closing time $T_{\rm c}$ can be calculated from the hodograph solution
\eqref{eq:13}
\begin{equation}
  \label{eq:50}
  T_{\rm c} = -3 \frac{1+r_0^2}{8 r_0^3} = \frac{3(1-2\sqrt{a_0}+2
    a_0)}{4(2\sqrt{a_0}-1)^3} .
\end{equation}
The corresponding soliton slope is zero by reflection symmetry and the
amplitude can be read off from $s = -r_0$, giving
\begin{equation}
  \label{eq:51}
  \begin{split}
    a(Y_{\rm c},T_{\rm c}) &= a_{\rm c} = (2\sqrt{a_0}-1)^2, \quad q(Y_{\rm c},T_{\rm
      c}) = q_{\rm c} = 0 . 
  \end{split}
\end{equation}
From this closing point, a constant region emerges, bounded by the
edges of the simple wave \eqref{eq:45} and its symmetric reflection
\begin{equation}
  \label{eq:52}
  |Y| = V(r_0,-r_0)(T-T_{\rm c}) = -\frac{2}{3} r_0 (T-T_{\rm c}) =
  \frac{2}{3}(2\sqrt{a_0}-1)(T-T_{\rm c}) .
\end{equation}
This constant region corresponds to the emergence of a line soliton
with amplitude $0 < a_{\rm c} < 1$.  Our findings for the bent-stem
soliton are summarised in Table~\ref{tab:bent_stem}.

\begin{table}
  \centering
  \begin{tabular}{c|cc}
    simple wave interaction region & unbounded & bounded \\[3mm]
    constraints on initial data & $0 \le a_0 \le
    \frac{1}{4} \iff \frac{1}{2} \le q_0 \le 1$ & 
    $\frac{1}{4} < a_0 < 1 \iff 0 < q_0 < \frac{1}{2}$ \\[3mm]
    initial geometric constraints & strongly bent soliton ($q_0^2 >
    a_0$) & weakly bent 
    soliton ($q_0^2 < a_0$) \\[3mm]
    long time dynamics & decaying parabolic soliton & non-decaying
    line soliton
  \end{tabular}
  \caption{Dynamics of the bent-stem soliton initial data
    \eqref{eq:KinkIC}. }
  \label{tab:bent_stem}
\end{table}

\begin{figure}
  \centering
  \includegraphics[trim={0 1cm 0 0},clip]{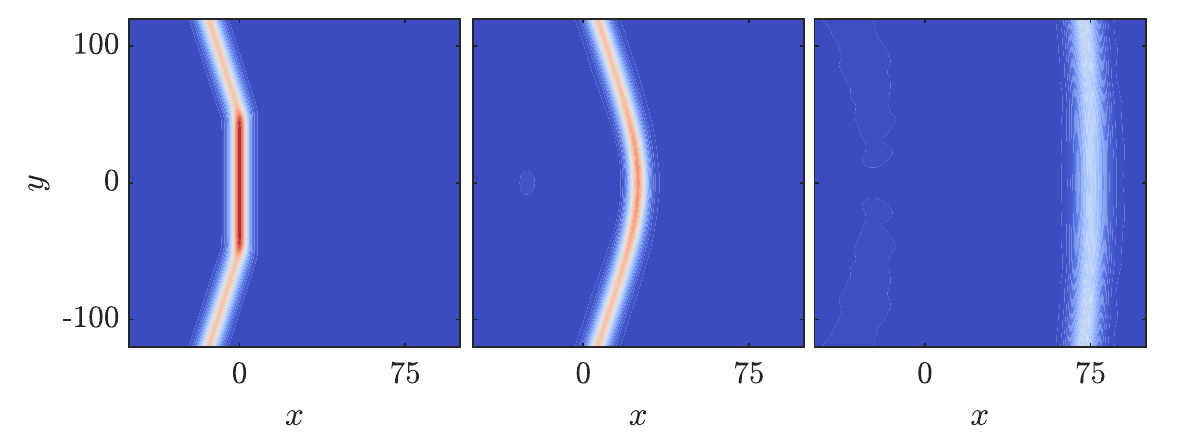}
 \includegraphics[]{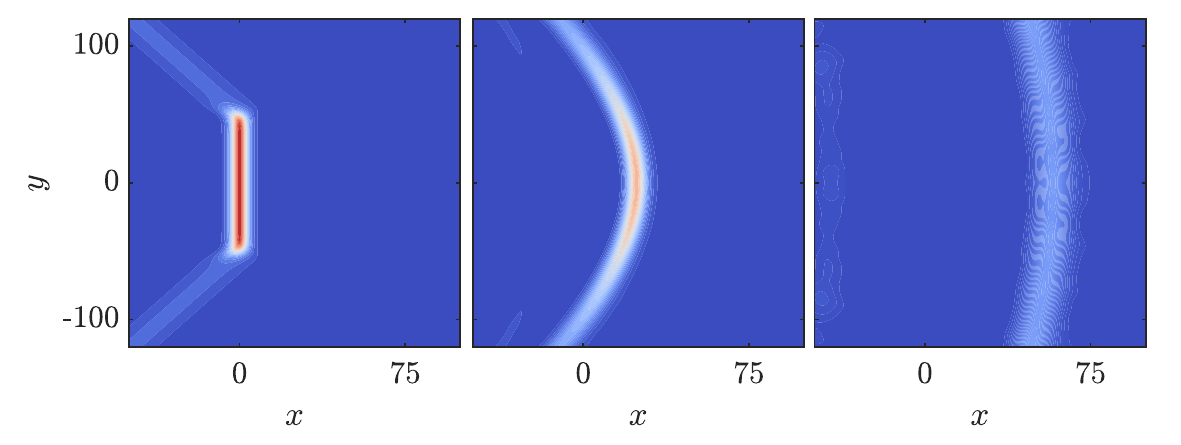}
 \caption{Top: numerical simulation of the bent-stem soliton for
   $\sqrt{a_0}=0.8$ and $q_0=0.2$ with $t \in (0,80,400)$ showing the
   emergence of a straight line soliton. The interaction region
   closing time is predicted to be $t = \ell T_{\rm c} = 236$
   (cf.~Fig.~\ref{fig:InteractionRegion} right). Bottom: numerical
   simulation of the bent-stem soliton for $\sqrt{a_0}=0.3$ and
   $q_0=0.7$ with $t \in (0,80,400)$ leading to a decaying parabolic
   soliton.  In both cases, $\ell = 100$.}
  \label{fig:bent_stem}
\end{figure}

\begin{figure}
  \centering
  \includegraphics[scale=.25]{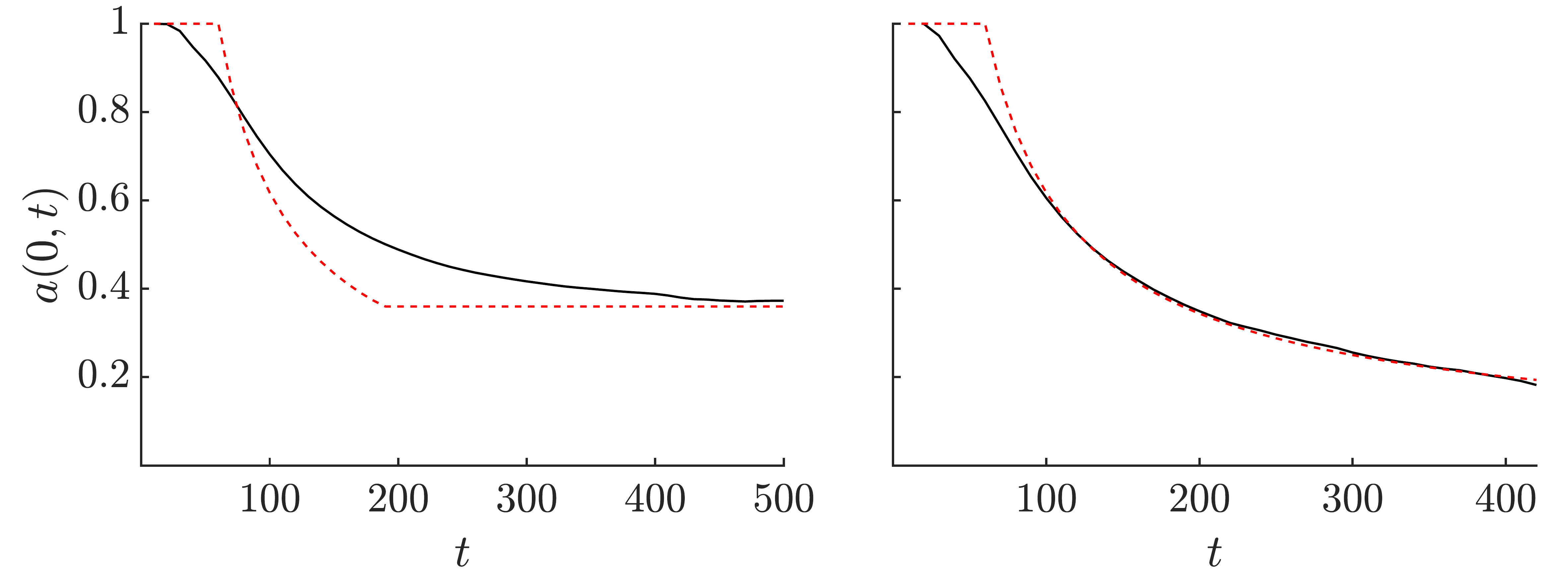}
  \caption{Comparison between numerical simulation (solid line) and
    the modulation solution (dashed line) of the bent-stem amplitude
    decay at $y=0$ for the parameters in Fig.~\ref{fig:bent_stem} top
    (left) and bottom (right). In order to account for the smooth
    initial data, the modulation solution is ``fitted'' by choosing
    $\ell=80$.}
  \label{fig:bent_stem_accuracy}
\end{figure}

Figure \ref{fig:bent_stem} depicts the numerical evolution of
bent-stem solitons for each of the scenarios in Table
\ref{tab:bent_stem}.  For the panels in Figure \ref{fig:bent_stem}
top, the initial conditions are nominally $\sqrt{a_0}=0.8$ and
$q_0=0.2$, with $\ell=100$. From our analysis, the emergence of a
constant region in the modulation, i.e., a vertical soliton with
amplitude $a_c = 0.36$, should begin to appear at $t = \ell T_c
\approx 236$. By $t=400$ in Fig.~\ref{fig:bent_stem} top, right, the
vertical soliton has emerged with amplitude very close to the
predicted value $0.36$ shown in Fig.~\ref{fig:bent_stem_accuracy}
left. In contrast, for the panels in Figure \ref{fig:bent_stem}
bottom, the initial conditions are $\sqrt{a_0}=0.3$ and $q_0=0.7$,
again with $\ell=100$. As expected, the system forms a parabola which
slowly decays over time.

For quantitative analysis, we consider the amplitude decay at $y=0$
for the bent stem simulations in Figure
\ref{fig:bent_stem_accuracy}. On the left is displayed the amplitude
decay for the weakly bent simulation shown in Fig.~\ref{fig:bent_stem}
top, while on the right is data from the strongly bent simulation from
Fig.~\ref{fig:bent_stem} bottom. Here we observe some deviation of the
numerical simulation from modulation theory for shorter times.  We
attribute these differences to higher order dispersive effects that
are not captured by the leading order modulation equations
\eqref{eq:modulationEqs}. However, the large $t$ predicted behaviour
agrees quantitatively with numerical simulations. For the weakly bent
stem, the amplitude asymptotically approaches $a_c=0.36$ as predicted,
while for the strongly bent stem case, the amplitude continues to
decrease as $t \rightarrow \infty$. As in the truncated case, we
slightly reduce the length $\ell$ in the modulation solution to
$\ell=80$ in order to account for the smoothing of the initial
conditions.

\begin{figure}
  \centering
  \includegraphics[trim={0 1cm 0
    0},clip]{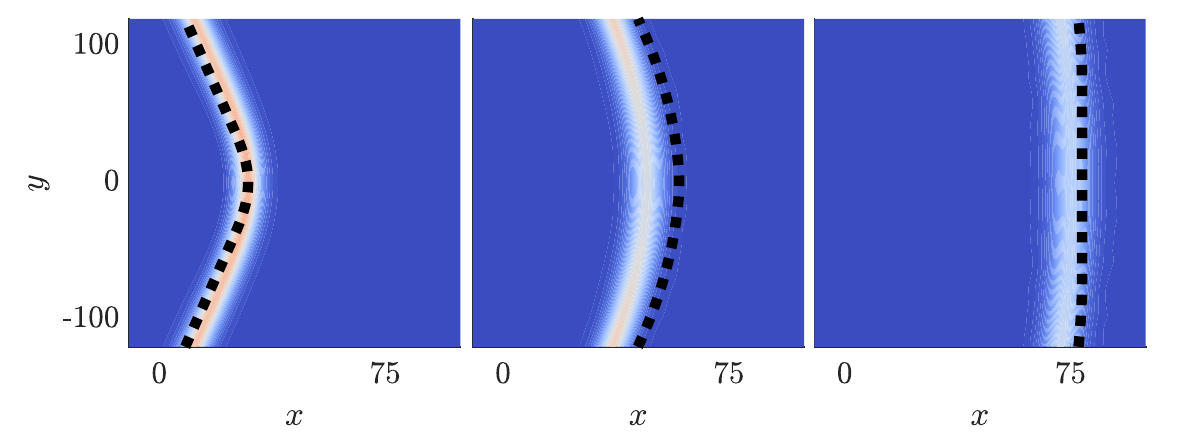}
  \includegraphics{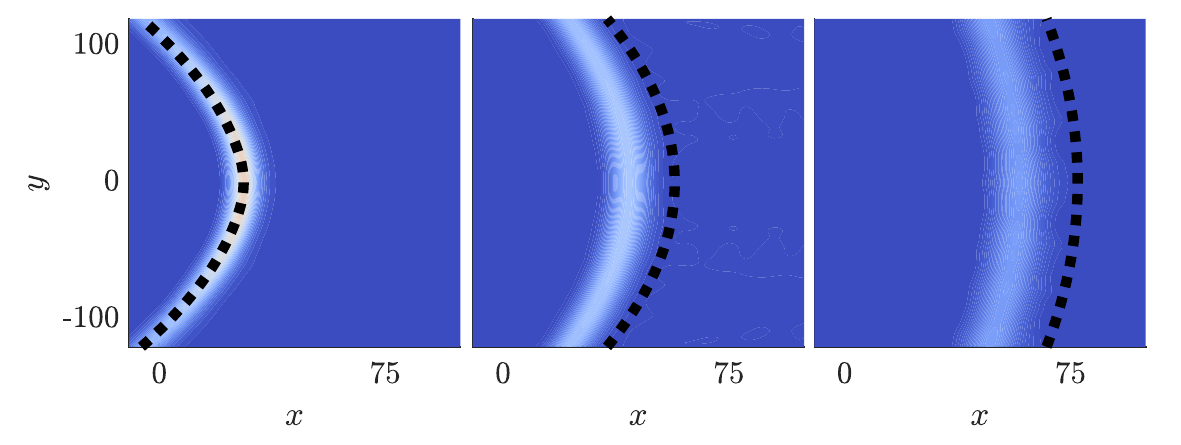} 
  \caption{Modulation solution phase (dashed) overlaid on contour
    plots for weakly (top) and strongly (bottom) bent-stem initial
    conditions when $t \in (100,200,400)$. The initial conditions are
    the same as in Figure \ref{fig:bent_stem}.}
  \label{fig:bent_stem_phase}
\end{figure}

We also consider the predicted soliton phase $\xi$ compared to the
numerical simulations. This is shown in Figure
\ref{fig:bent_stem_phase}. The overlaid predicted phases (dashed
curves) were generated by using the modulation solution for $q$ and
then numerically integrating for $\xi$ according to
Eq.~\eqref{eq:39}. We utilised the speed $c(y,t)$ in the prediction
after fitting the phase so that it lines up with the front's maximum
along $y = 0$ at $t = 100$ in the leftmost panels.  The ensuing phase
profiles at $t = 200$ and $t = 400$ are slightly advanced relative to
the numerical simulation, which can be attributed to higher order
phase errors that are common in soliton perturbation theory.  Such a
correction would result in an additional term $\xi_0(x,y,t)$ being
added to the modulated soliton phase $\xi$ in Eq.~(\ref{eq:39}).
Importantly, the shape of the front's crest is well-described by the
modulation solution for $q$.

It is evident that both the weakly and strongly bent-stem numerical
evolutions are well approximated by the modulation solution,
asymptoting to a line soliton and a parabolic wave in long time,
respectively.

\subsection{Bent solitons}
\label{sec:kink-modulation}

We now consider the bent-stem soliton initial data \eqref{eq:KinkIC}
with a vanishing stem $\ell \to 0$, i.e., a bent soliton in which
\begin{equation}
    \label{eq:ArrowheadIC}
    a(y,0)=a_0, \quad
    q(y,0)=
    \begin{cases} 
        q_0 & y>0 \\
        -q_0 & y\le 0 
    \end{cases} .
\end{equation}
Figure \ref{fig:initial_data} right displays the corresponding initial
condition for the KPII equation \eqref{eq:KP}. In contrast to the
truncated and bent-stem soliton initial conditions, the initial
conditions (\ref{eq:ArrowheadIC}) for the modulation equations
\eqref{eq:modulationEqs} correspond to a Riemann problem.  We limit
our consideration to an expansive Riemann problem by taking $q_0>0$.
This case corresponds to a partial soliton interacting with an
expansive corner (cf.~Fig.~\ref{fig:mach_expansion}).  If $q_0 < 0$, a
case we do not consider, the partial soliton interacts with a
compressive corner and gives rise to regular and Mach reflection
(cf.~Fig.~\ref{fig:mach_reflection}).

For the case where $q_0 + \sqrt{a_0} = 1$, cf.~\eqref{eq:26}, the bent
soliton's evolution can be obtained directly from the bent-stem
soliton evolution by taking the vanishing stem limit $\ell \to
0$. Consequently, the bent soliton inherits the bent-stem soliton's
bifurcation in long-time dynamics.  When $\sqrt{a_0} > q_0$, the
modulation solution for the bent-stem soliton post simple wave
interaction ($T > T_{\rm c}$) exhibits an expanding constant region
with $a = a_{\rm c}$, $q = q_{\rm c}$ in Eq.~\eqref{eq:51} bounded by
the characteristics \eqref{eq:52}. We refer to this as the strong
interaction case.

When $\sqrt{a_0}<q_0$ and $q_0 + \sqrt{a_0} = 1$, we can take the
$\ell \to 0$ limit of the bent-stem soliton with an unbounded
interaction region. In this limit, the corners of the interaction
region in Eq.~\eqref{eq:48} are $(\pm y_*,t_*) = (\pm \ell Y_*,\ell
T_*) \to (0,0)$.  Additionally, the interaction boundaries
\eqref{eq:7} and \eqref{eq:49} collapse to $y = 0$.  However, the
characteristics \eqref{eq:45} leaving the boundaries of the
interaction region persist.  At $y = 0$, the soliton amplitude in the
interaction region is explicitly \eqref{eq:12} so that $a(0,t) \to 0$
as $\ell \to 0$.  This case corresponds to weak interaction.

In order to elucidate more details and provide an alternative method
of solution, we now solve the Riemann problem \eqref{eq:ArrowheadIC}
for the bent soliton directly.  We relax the assumption $q_0 +
\sqrt{a_0} = 1$ and consider general $a_0 > 0$ and $q_0 > 0$, which is
equivalent to applying the scaling symmetry (\ref{eq:31}) to the
bent-stem soliton problem and taking $\ell \to 0$.

First, we consider the strong interaction case $\sqrt{a_0}>q_0$.  The
upper and lower solitons cannot be connected by a single simple wave,
which would require $\sqrt{a_0}-q_0=\sqrt{a_0}+q_0$, leading to the
conclusion $q_0 = 0$.  Instead, we introduce the intermediate state
$(a_{\rm i},q_{\rm i})$ and connect it to $(a_0,\pm q_0)$ with simple
waves satisfying
\begin{equation}
  \label{eq:4}
  \begin{split}
    \sqrt{a_{\rm i}}-q_{\rm i} = \sqrt{a_0}-q_0 , \quad
    \sqrt{a_{\rm i}}+q_{\rm i} = \sqrt{a_0}+(-q_0) .
  \end{split}
\end{equation}
The solution to these equations under the given constraints is 
\begin{equation}
  \label{eq:63}
  q_{\rm i}=0, \quad \sqrt{a_{\rm i}}=\sqrt{a_0}-q_0,
\end{equation}
where we use a 2-RW to connect the intermediate state to the top
soliton and a 1-RW to connect to the bottom soliton.

\begin{figure}
  \centering
  \includegraphics{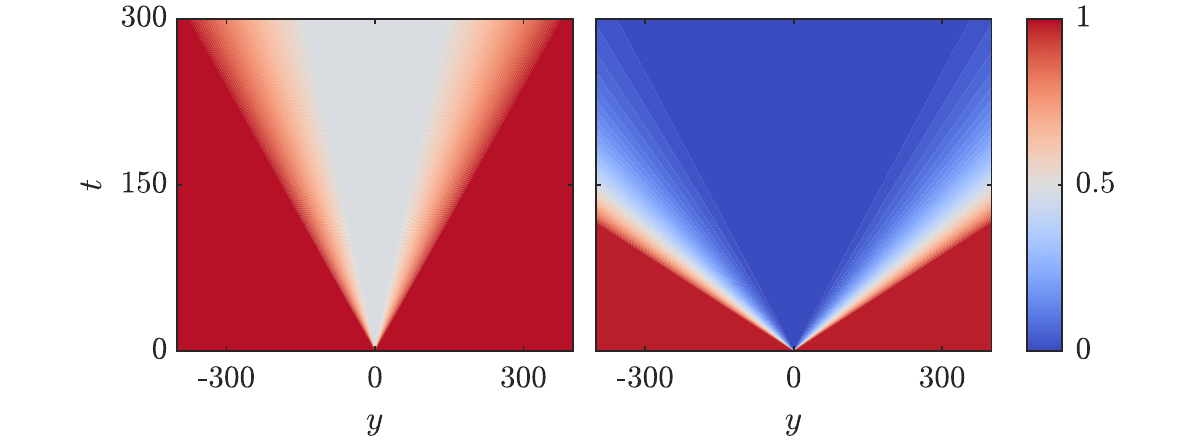}
  \caption{Space-time contour plots of amplitude modulation solutions
    for bent soliton initial data.  Left: The strong interaction case
    \eqref{eq:4} with $a_0 = 1 > q_0 = 0.3$.  Right: The weak
    interaction case \eqref{eq:53} with $a_0 = 1 < q_0 = 1.4$. The two
    cases are similar but note the nonzero central amplitude for
    strong interaction.}
  \label{fig:bent_amp_contour}
\end{figure}
Consequently, evolving the bent soliton over time gives an
intermediate constant region connected to the two canted solitons by
centred simple waves
\begin{equation}
  \label{eq:5}
  \begin{split}
    \sqrt{a(y,t)} &=
    \begin{cases}
      \sqrt{a_0} & V(a_0,q_0)t < |y| \\
      \frac{3}{8}(\frac{y}{t}+2\sqrt{a_{\rm i}}) & ~\,V(a_{\rm i},0)t
      < |y| < V(a_0,q_0)t \\
      \sqrt{a_{\rm i}} & \qquad \qquad \quad ~\, |y| < V(a_{\rm i},0)t
    \end{cases},\\
    q(y,t) &=
    \mathrm{sgn}(y) \begin{cases}
      q_0 & V(a_0,q_0)t < |y| \\
      \sqrt{a(y,t)}-\sqrt{a_{\rm i}} &
      ~\,V(a_{\rm i},0)t < |y| < V(a_0,q_0)t \\
      0 & \qquad \qquad \quad ~\, |y| < V(a_{\rm i},0)t
    \end{cases} 
  \end{split},
\end{equation}
where $V(a,q) = 2q + \frac{2}{3} \sqrt{a}$.  The solution contains a
vertical line soliton expanding in $y$ with amplitude $a_{\rm i}$ in
Eq.~(\ref{eq:63}).  This solution agrees with our analysis of the
$\ell \to 0$ limit of the bent-stem soliton when
$q_0 + \sqrt{a_0} = 1$.  As we will show in the next subsection, this
strong interaction case corresponds to Mach expansion of a soliton
interacting with a corner.

\begin{figure}
  \centering
  \includegraphics{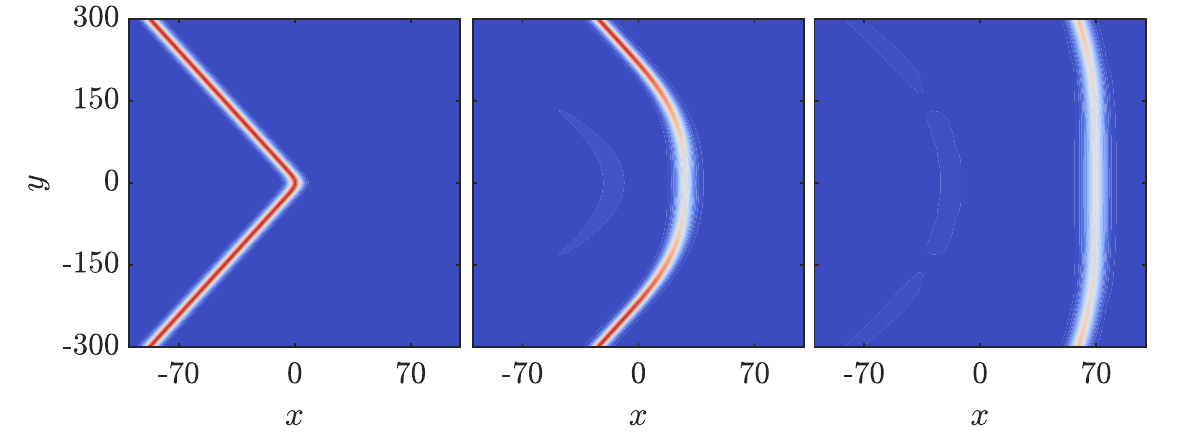}
  \includegraphics{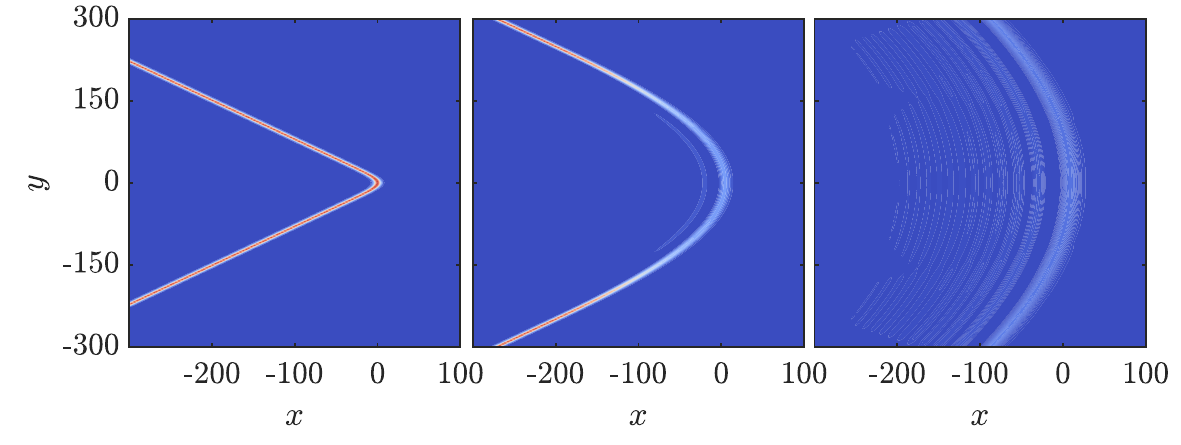}
  \caption{Numerical simulation of bent solitons for the strong
    interaction when $a_0=1$, $q_0=0.3$ for $t \in (0,150,400)$ in the
    top panels and the weak interaction when $a_0=1$, $q_0=1.4$ when
    $t \in (0,60,220)$ in the bottom panels. }
  \label{fig:bent}
\end{figure}
For weak interaction when $\sqrt{a_0}<q_0$, the above calculation
fails because $\sqrt{a_{\rm i}}$ exhibits negative values.  This
determines the critical slope
\begin{equation}
  \label{eq:65}
  q_{\rm cr} = \sqrt{a_0}
\end{equation}
between the two classes of bent soliton dynamics.  Instead, we
introduce the intermediate vacuum state $a_{\rm i} = 0$. In order to
connect to vacuum with a simple wave from each of the bent solitons,
we require $q_{\rm i} \ne 0$. Since $q$ in the vacuum region is
undefined, we determine it locally based on the simple wave criterion.
By symmetry,
\begin{equation}
  \label{eq:lim}
  q_{\rm i+} = \lim_{y \to 0^+} q(y,t)=-\lim_{y \to 0^-}q(y,t) =q_{\rm i-},
\end{equation}
with the initial discontinuity at $y=0$. We can use Riemann invariants
to calculate the values of $q_{\rm i \pm}$. For the top simple wave,
$\sqrt{a_0}-q_0=-q_{\rm i +}$, and by symmetry $q_{\rm i +}=-q_{\rm i -}$. The
solution for the top simple wave is
\begin{equation}
  \label{eq:53}
  \begin{split}
    \sqrt{a(y,t)} &=
    \begin{cases}
      \sqrt{a_0} & \, V(a_0,q_0)t < |y| \\
      \frac{3}{8}(\frac{y}{t}-2q_{\rm i +}) & V(0,q_{\rm i +})t
      < |y| < V(a_0,q_0)t \\
      0 & \qquad \qquad \qquad \! |y| < V(0,q_{\rm i +})t
    \end{cases},\\
    q(y,t) &=
    \mathrm{sgn}(y) \begin{cases}
      q_0 & \, V(a_0,q_0)t < |y| \\
      \sqrt{a(y,t)}+q_{\rm i +} &
      V(0,q_{\rm i +})t < |y| < V(a_0,q_0)t \\
      q_{\rm i +} & \qquad\qquad\qquad \! |y| < V(0,q_{\rm i +})t
    \end{cases} 
  \end{split},
\end{equation}
with a symmetric reflection for the bottom simple wave. This solution
consists of an expanding vacuum $a = 0$ region connected to outgoing,
canted line solitons by simple waves. These simple waves are rotated
versions (see Eq.~\eqref{eq:34}) of the partial soliton solution
\eqref{eq:38}.  They are completely disconnected from one another; at
this order of approximation, the interaction is negligible in that the
evolution of the upper and lower branches can be analysed
independently of one another.  The vacuum region's rate of expansion
is proportional to $q_0-\sqrt{a_0} > 0$. A more ``bent'' initial
soliton ($q_0 > q_{\rm cr}$) causes the outgoing solitons to separate
from one another sufficiently fast so that their interaction is
negligible within the context of modulation theory.  As we will show
in the next subsection, this weak interaction case corresponds to
regular expansion of a soliton interacting with a corner.

The numerical simulations in Figure \ref{fig:bent} are essentially
consistent with these predictions.  For the bent soliton with $a_0=1$
and $q_0=1.4 > q_{\rm cr} = 1$ in Figure \ref{fig:bent} bottom, a
decaying parabolic front with trailing oscillations appear.  Although
an expanding, strictly vacuum region predicted by modulation theory is
not immediately apparent, amplitude decay is present.  We have
verified that the amplitude, shape, and propagation of the leading
parabolic front is consistent with the profile for the cKdV parabolic
soliton \eqref{eq:11}.  The trailing oscillations are consistent with
a two-dimensional generalisation of the oscillatory shelf that is
common for perturbed KdV (cKdV) problems.  In contrast, for an initial
bent soliton with $a_0=1$ and $q_0=0.3 < q_{\rm cr} = 1$, as seen in
Figure \ref{fig:bent} top, a new vertical line soliton with reduced
amplitude appears.

\begin{figure}
  \centering
    \includegraphics[scale=.25]{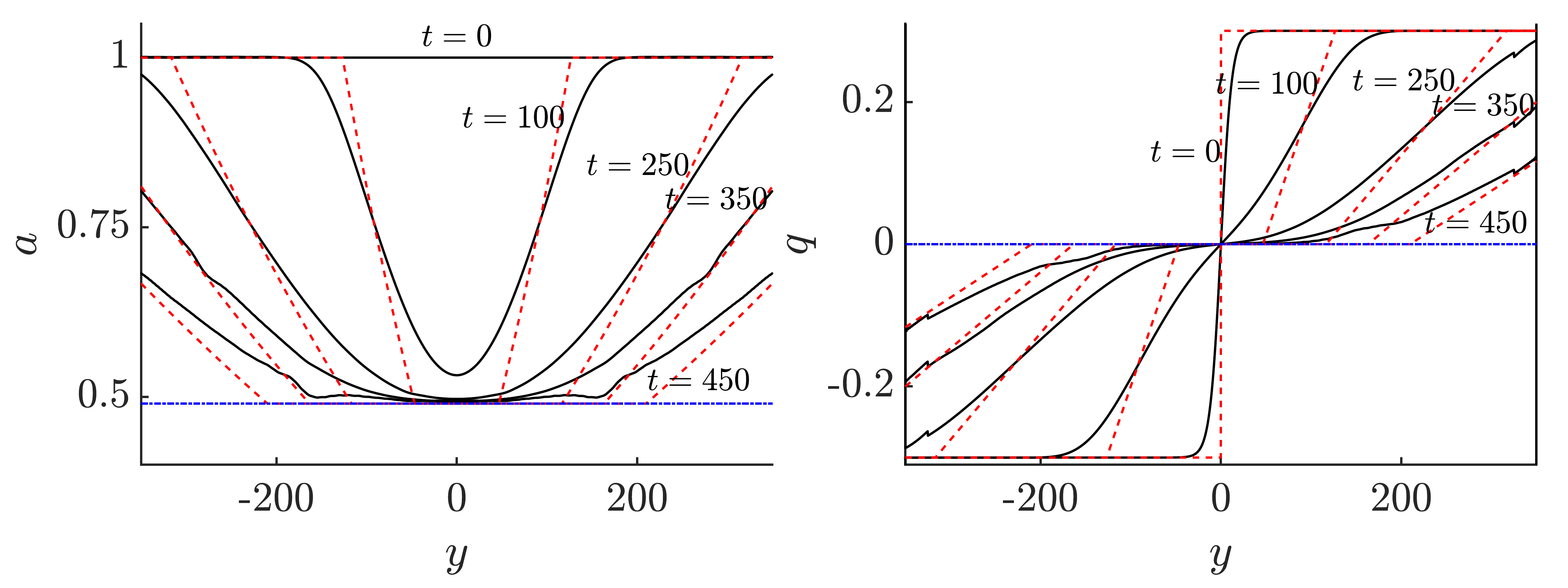}
    \caption{Modulated soliton amplitude $a$ and slope $q$ extracted
      from the numerical simulation in the strong interaction case of
      Fig.~\ref{fig:bent} top (solid curves) and the modulation
      solution (\ref{eq:5}) (dashed curves) at different times.}
  \label{fig:bent_accuracy}
\end{figure}

Quantitative results further confirm our analysis. In Figure
\ref{fig:bent_accuracy}, it is evident that the predicted solution for
both soliton amplitude and slope, extracted according to
Eq.~\eqref{eq:60} captures the behaviour of the strong interaction
case with $a_0=1$ and $q_0=0.3 < q_{\rm cr} = 1$. As expected, the
solution for large times approaches a line soliton with
$a_{\rm i} \approx 0.49$.

\begin{figure}
  \centering
  \includegraphics[scale=.25]{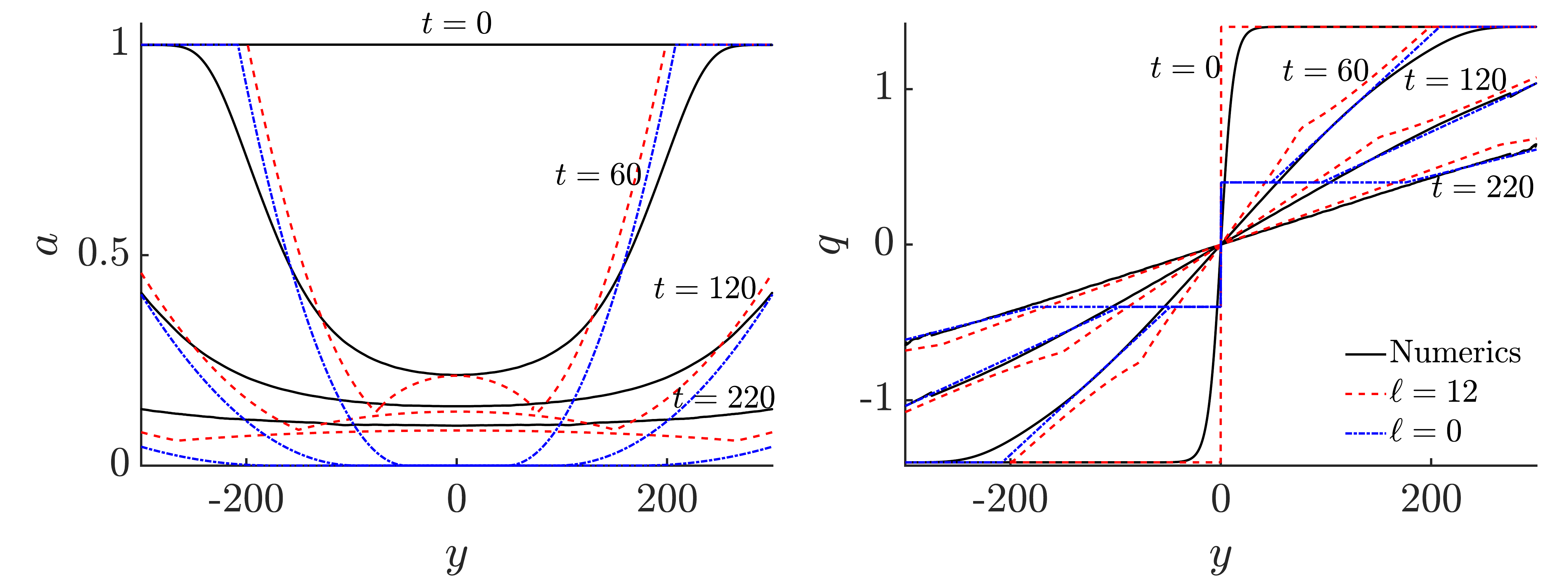}
  \caption{Comparison of bent-stem soliton modulation with small stem
    $\ell = 12$ (dashed, red) and no stem $\ell \to 0$ (dash-dotted,
    blue) to numerical simulation (solid, black) of the weakly
    interacting bent soliton evolution in Fig.~\ref{fig:bent} bottom.}
  \label{fig:bent_stem_compare_bent}
\end{figure}

For the weak interaction case $q_0 = 1.4 > q_{\rm cr} = 1$ shown in
Fig.~\ref{fig:bent_stem_compare_bent} ($\ell = 0$ case), the
simulation's lead wave slope (solid) is well approximated by the
modulation solution (dash-dotted).  The amplitude does not reach zero
as modulation theory predicts, although it does continually
decrease. Recalling that modulation theory applies under slowly
varying assumptions, it is not surprising that an immediate transition
from unit amplitude to zero amplitude does not occur in the numerical
simulation.  In Figures \ref{fig:bent_stem_accuracy} and
\ref{fig:bent_accuracy}, we observe that the numerical solution
temporally lags behind the modulation solution.  The same happens here
in Fig.~\ref{fig:bent_stem_compare_bent}, albeit to a more significant
degree in amplitude.  While this weakly interacting bent soliton
simulation deviates from the modulation solution in amplitude, in fact
it can be reasonably approximated by the bent-stem modulation solution
for moderate stem length $\ell$. This is to be expected; the bent-stem
analysis for sufficiently large $q_0$ shows that any positive stem
length $\ell>0$ implies algebraic amplitude decay to zero
(cf.~Eq.~\eqref{eq:9}) rather than a sudden amplitude decrease to zero
in finite time.  The initial smoothing of the numerical simulations
(see Appendix~\ref{sec:numerical-method-1}) can be viewed as an
effective $\ell>0$ for these bent soliton simulations.  Consequently,
we compare the bent-stem soliton modulation solution for the initial
data \eqref{eq:KinkIC} (rescaled according to \eqref{eq:31} so that
the canted, outgoing solitons have unit amplitude) with $\ell = 12$ to
the numerical simulation of the weakly interacting bent soliton in
Fig.~\ref{fig:bent_stem_compare_bent}.  Now the decaying parabolic
front is represented in the modulation solution.  In other words, the
weakly interacting bent soliton evolution exhibits a remnant of the
bent-stem soliton solution.

\subsection{Regular and Mach expansion}
\label{sec:regul-mach-expans}

We are now in a position to interpret this analysis in the context of
the soliton-corner initial boundary value problem that is
schematically depicted in Fig.~\ref{fig:mach_expansion}.  Consider a
vertical, partial soliton with amplitude $a$ propagating in the
positive $x$ direction adjacent to a horizontal wall located at
$y = 0$.  When this soliton encounters a corner at the origin that
suddenly opens or turns away by the clockwise angle $\varphi > 0$, it
expands.  The nature of its expansion depends on the corner angle and
soliton amplitude.  Via the nonlinear method of images, we map the
partial soliton at the moment it encounters the corner to the bent
soliton initial data for the numerical simulation in
Fig.~\ref{fig:initial_data} right and for modulation theory in
Eq.~(\ref{eq:ArrowheadIC}) with $a_0 = a$ and $q_0 = \tan{\varphi}$.
Consequently, the critical corner angle separating two distinct types
of soliton-corner interaction is (cf.~Eq.~\eqref{eq:65})
\begin{equation}
  \label{eq:66}
  \tan{\varphi_{\rm cr}} = \sqrt{a}.
\end{equation}

For the case $\varphi > \varphi_{\rm cr} = \arctan{\sqrt{a}}$ (sharp
corner), the soliton almost completely separates from the wall.  The
residual soliton-wall interaction is through a decaying parabolic
soliton.  In turn, the propagating partial line soliton decays,
retreating further away from the wall.  We term this case
\textit{regular expansion}.  In contrast, for slight turns of the wall
at the corner where
$0 < \varphi < \varphi_{\rm cr} = \arctan{\sqrt{a}}$, the soliton also
develops a curved front that instead terminates at a non-decaying
soliton perpendicular to the wall with lower amplitude than the
incident soliton.  The predicted soliton wall amplitude is
\begin{equation}
  \label{eq:74}
  a_{\rm w} = (\sqrt{a}-\tan{\varphi})^2.
\end{equation}
Despite propagating away from the wall, it does not ``escape'' its
influence like in the regular expansion case.  The residual soliton
formed at the wall is the expansion counterpart to the Mach stem that
forms during the course of Mach reflection
(cf.~Fig.~\ref{fig:mach_reflection}).  We term this case \textit{Mach
  expansion}.  Surprisingly, the crossover from regular to Mach
expansion occurs at precisely the same corner angle \eqref{eq:66} as
the crossover from regular to Mach reflection \eqref{eq:64}. These
results are highlighted in Table \ref{tab:corner}.

\begin{table}
  \centering
  \begin{tabular}{c|cc}
    soliton-corner expansion & regular expansion & Mach expansion \\[3mm]
    soliton amplitude $a$, corner angle $\varphi$ & $0 < \sqrt{a} <
    \tan{\varphi}$ & $\tan{\varphi} \le \sqrt{a}$ 
     \\[3mm]
    long time dynamics at wall & decaying parabolic soliton &
    non-decaying line soliton
  \end{tabular}
  \caption{Soliton-expansive corner initial, boundary value problem.}
  \label{tab:corner}
\end{table}

\section{Discussion and conclusion}
\label{sec:conclusion}

Using the KPII equation as a model of multidimensional gravity wave
solitons, we describe the evolution of truncated and interacting
oblique solitons using modulation theory, and we compare the
analytical predictions with the results from direct numerical
simulations.  The initial value problems considered are distinguished
by geometric configurations of partial solitons that propagate away
from one another.  Despite this, residual interactions between
solitons occur that lead to nontrivial wave patterns.

An initial soliton that is transversely confined or truncated, and
that propagates into an open region, first ``curls'' at the endpoints
as it then morphs into a parabolic shape over long time.  The front's
parabolic shape flattens with a linearly increasing focal length with
time.  The parabolic wave's amplitude and speed exhibits algebraic
decay proportional to $t^{-2/3}$.  Such wave patterns appear to be
common in images of oceanic internal waves for near-shore conditions
\cite{jackson_atlas_2004,pan_analyses_2007,wang_internal_2017}.

We also generalise the above truncated soliton configuration by
appending canted partial solitons to it and find new dynamical
behaviour.  In addition to the decaying parabolic wavefront for
sufficiently canted solitons, a non-decaying vertical soliton with
reduced amplitude relative to the original soliton segment appears.
This bifurcation in behaviour carries over to the soliton-corner
expansion problem.

The final initial value problem for a bent soliton also describes the
interaction of a soliton propagating parallel to a wall with an
expansive corner.  For a sharp enough bend, the solitons exhibit weak
interaction through a decaying parabolic front, the case we identify
as regular expansion of a soliton.  For a slight enough bend, the
solitons continue to interact so as to produce an intermediate,
non-decaying soliton at the wall with reduced amplitude $a_{\rm w}$
\eqref{eq:74} that connects them.  This case of Mach expansion
parallels the well-known Mach reflection of oblique solitons, both of
which occur at the same critical angle and display a similar
transition between strong and weak interactions.  Such a transition in
the wave dynamics could potentially be observed in the shallow water
context by soliton generation from a moving disturbance
\cite{lee_upstreamadvancing_1990,li_three-dimensional_2002} or by
experiments analogous to previous shallow water studies involving Mach
reflection
\cite{perroud_solitary_1957,melville_mach_1980,li_mach_2011,kodama_kp_2016}.

In the Mach reflection case, an important quantitative test of the
theory is its prediction of amplitude amplification of the Mach stem
at the wall.  After properly taking into account a higher-order
asymptotic approximation to shallow water waves than the KP
equation, the soliton wall amplitude has been demonstrated to
satisfactorily predict experiments across a range of incident
soliton parameters (angle and amplitude) \cite{kodama_kp_2016}.
This points to a possible quantitative test of the Mach expansion
theory presented here by measuring the wall reduction factor
\begin{equation}
  \label{eq:75}
  \alpha \equiv \frac{a_{\rm w}}{a} = \left ( 1 -
    \frac{\tan{\varphi}}{\sqrt{a}} \right )^2, \quad \tan{\varphi} < \sqrt{a}
\end{equation}
in similar shallow water experiments.  The factor $\alpha < 1$ is the
ratio of wall soliton amplitudes post ($a_{\rm w}$) and pre ($a$)
corner interaction, respectively, predicted by KP theory.  We note
that, in order to achieve quantitative agreement with experiment, it
may be necessary to incorporate higher order effects such as those
considered in \cite{kodama_kp_2016}.

These results also motivate the conjecture that outgoing gravity line
solitons propagating away from one another with slopes $\pm q_\infty$
and similar amplitudes $a_\infty$ lead to a decaying parabolic or
negligible interaction region when sufficiently sloped $q_\infty \ge
\sqrt{a_\infty}$ but leave a residual line soliton between them when
$q_\infty < \sqrt{a_\infty}$.

This work demonstrates the practical utility and efficacy of soliton
modulation theory to describe rich nonlinear wave dynamics.  All the
solutions that we consider are globally existing simple wave or
interacting simple wave solutions of the hyperbolic modulation
equations.  These solutions, when projected back onto a line soliton,
quantitatively agree with direct numerical simulations of the KPII
equation.  Although not previously recognised as such, simple
wave-modulated solitons can also be seen in a variety of previous KPII
numerical studies
\cite{funakoshi_reflection_1980,kodama_soliton_2009,kao_numerical_2012,chakravarty_numerics_2017}.
In particular, the transient portion of the reflected wave that
develops during both regular and Mach reflection of a soliton by a
corner appears to show a similar wave pattern to the partial soliton
studied here.  An intriguing problem is to consider the modulation
equations with initial data that is compressive, i.e., that would give
rise to shock solutions.  Indeed, two colliding partial solitons and
``V-shaped'' initial conditions
\cite{kodama_soliton_2009,chakravarty_numerics_2017} for regular and
Mach reflection, give rise to compressive Riemann problems for the
modulation equations (\ref{eq:modulationEqs}).  How are such initial
value problems regularised?  What do shocks mean in this modulation
context?

\section*{Acknowledgements}

The work of MAH and SR was supported by NSF grant DMS-1816934.  The
work of MM was supported by the NSF GRFP.  The work of GB was
supported by NSF grant DMS-2009487.  Authors thank the Fields
Institute Focus Program on Nonlinear Dispersive Partial Differential
Equations and Inverse Scattering in the summer of 2017 where this
research was initiated.

\appendix 

\section{Derivation of the soliton modulation equations}
\label{sec:deriv-modul-equat}

Here we show how Eqs.~\eqref{eq:modulationEqs} can be directly derived
from the KP equation \eqref{eq:KP} using multiple scales, without
employing the full Whitham modulation theory.  First, we introduce the
rescaling
\begin{equation}
  \label{eq:54}
  X = \epsilon x, \quad Y = \epsilon y, \quad T = \epsilon t
\end{equation}
into the KPII equation \eqref{eq:KP}
\begin{equation}
  \label{eq:55}
  \left(u_T + uu_X + \epsilon^2 u_{XXX}\right)_X +u_{YY} = 0.
\end{equation}
In order to study modulated line solitons, the asymptotic expansion
\begin{equation}
  \label{eq:56}
  \begin{split}
    u(X,Y,T;\epsilon) &= u_0(\xi,Y,T) + \epsilon u_1(\xi,Y,T) +
    \cdots, \quad u_0(\xi,Y,T) = a(Y,T) \mathrm{sech}^2 \left (
      \eta \xi \right
    ), \\
    \eta \xi_X &= \frac{1}{\epsilon} \sqrt{\frac{a(Y,T)}{12}}, \quad \eta \xi_Y = \frac{1}{\epsilon}
    \sqrt{\frac{a(Y,T)}{12}} q, \quad \eta
    \xi_T = -\frac{1}{\epsilon}\sqrt{\frac{a(Y,T)}{12}} c(a,q) ,
  \end{split}
\end{equation}
is assumed where $\xi$ is the fast variable.  The coefficient
$\eta(Y,T) = \sqrt{a(Y,T)/12}$ is determined by the consistency
condition $\xi_{XY} = \xi_{YX} = 0$.  
The consistency condition $\xi_{YT} = \xi_{TY}$ yields the slope
modulation equation (\ref{eq:27}).  The amplitude modulation equation
(\ref{eq:25}) is obtained by inserting the ansatz (\ref{eq:56}) into
the KPII equation (\ref{eq:55}).  At first order in $\epsilon$, we
obtain an inhomogeneous ODE for $u_1$ that, when integrated once with
respect to $\xi$, is
\begin{equation}
  \label{eq:58}
  \begin{split}
    -c \partial_{\xi} u_1 + \partial_{\xi} \left ( u_0 u_1 \right )
    + \partial_{\xi\xi\xi} u_1 + q^2 \partial_{\xi} u_1 = - \left
      ( \partial_T u_0 + 2 q \partial_Y u_0 + q_Y u_0 \right ) .
  \end{split}
\end{equation}
Solvability over the space of $L^2(\R)$ solutions is enforced by
the orthogonality condition
\begin{equation}
  \label{eq:59}
  \left ( \partial_T + 2 q \partial_Y + q_Y
  \right ) \int_{\R} u_0^2 \, \mathrm{d}\xi = \left ( \partial_T + 2 q \partial_Y + q_Y
  \right ) \left (\frac{8\sqrt{3}}{3}
  a^{3/2}\right ) = 0,
\end{equation}
which, upon simplification, results in the amplitude modulation
equation (\ref{eq:25}).

\section{Numerical integration of the KP equation}
\label{sec:numerical-method-1}

To validate our analytical results, we implement the pseudospectral
method described in \cite{kao_numerical_2012}, which utilises a
hyper-Gaussian windowing function and Fourier discretisation for
non-periodic data in $y$ and periodic data in $x$. We essentially
follow \cite{kao_numerical_2012} with a few modifications. The method
proceeds as follows. Instead of solving the KP equation \eqref{eq:KP}
for the original function $u$, we instead solve an equivalent PDE for
a windowed function
\begin{equation}
    \label{window_func}
    v(x,y,t) = W(y)u(x,y,t),
\end{equation} where
\begin{equation}
  \label{eq:window}
  W(y) = e^{-a_n |y/L_y|^n},
\end{equation} 
with $n = 27$ and $a_n = 1.111^n\ln{10}$. The rapid decay of $W(y)$
for $|y|$ near $L_y$ ensures that $v=0$ near the top and bottom
boundaries of the domain so that its periodic extension is smooth. In
this region, we assume that the solution $u$ asymptotes to
non-modulated line solitons. Thus, $u$ can be decomposed as
\begin{equation}
    \label{eq:decomp}
    u = v+(1-W)\hat{u},
\end{equation}
where $\hat{u}$ are line solitons with constant $a$ and $q$ of the form shown in Eq.~\eqref{eq:KPsoliton}. Inserting the transformation \eqref{eq:decomp} into the KP equation \eqref{eq:KP}, we obtain an equivalent PDE for the windowed function $v$
\begin{subequations}
\label{eq:window_pde}
\begin{equation}
    \label{eq:window_pde_1}
    (v_t+vv_x+v_{xxx})_x +v_{yy} = (1-W)(W\hat{u}\hat{u}_x-(v\hat{u})_x+2W'\hat{u}_y+W''\hat{u})_x,
    \end{equation}
    subject to the initial conditions
    \begin{equation}
    \label{eq:window_pde_2}
    v(x,y,0) = u(x,y,0)-(1-W(y))\hat{u}(x,y,0).
\end{equation}
\end{subequations}
The above PDE \eqref{eq:window_pde} is solved numerically. At each
time $t$ we reconstruct the true solution $u$ using
Eq.~\eqref{eq:decomp}. The advantage of this method is that since $v$
is zero at the domain boundaries, we obtain spectral convergence using
a Fourier discretisation in space. In order to preserve spectral
accuracy, the derivatives $\hat{u}_x$, $\hat{u}_y$, $W'$, and $W''$ on
the right hand side of \eqref{eq:window_pde_1} are calculated
analytically; this is one difference from \cite{kao_numerical_2012}
where these derivatives are calculated using finite difference
approximations. Time stepping is performed with an integrating factor
and the classic fourth-order Runge-Kutta scheme. Simulations are
terminated before the windowing region is corrupted by non-solitonic
data.

\begin{figure}
  \centering
  \includegraphics[]{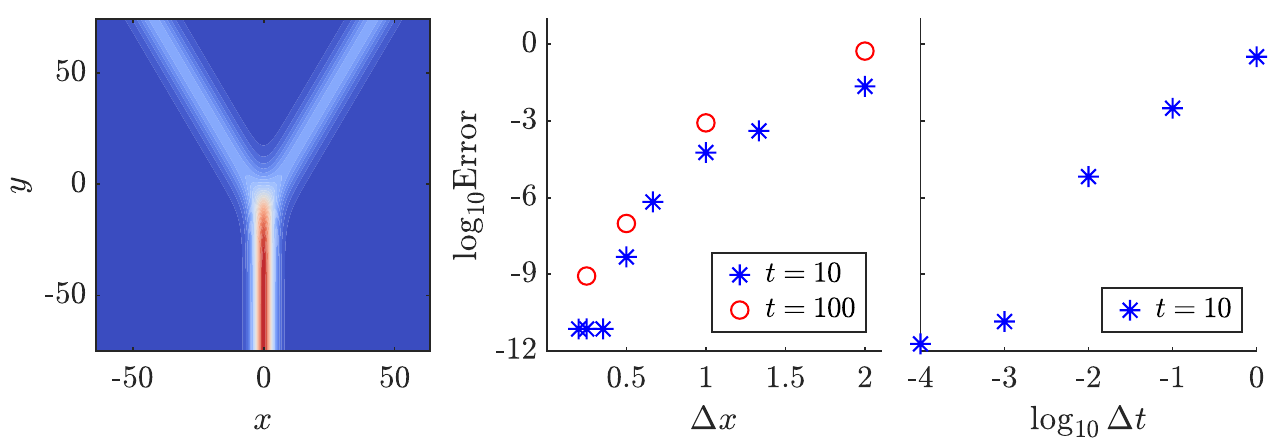}
  \caption{Convergence of Y-shaped soliton (left) in space (centre) and
    time. These simulations were run on the domain $[-128,128] \times
    [-64,64]$ with $\Delta x = \Delta y$. For the centre plot, we fixed
    $\Delta t = 10^{-3}$, and for the right we fixed $\Delta x = \Delta
    y = 1/4$.}
  \label{fig:convergence}
\end{figure}

The numerical scheme described above is validated using an exact
Y-shaped solution, also known as the Miles resonant soliton (see
Fig.~\ref{fig:convergence} left). By refining the time and space
steps, convergence is obtained at $t=10$ to approximately $10^{-12}$
in the 2-norm relative difference of the numerical and exact
travelling wave solutions, as shown in
Figure~\ref{fig:convergence}. The rate of spatial convergence in
Figure~\ref{fig:convergence} demonstrates that spectral accuracy is
obtained. In order to ensure reasonable computation time and memory
demands, the simulation parameters are fixed at the grid spacing
$\Delta x = 1/2$ and time step $\Delta t = 10^{-3}$. Based on
Figure~\ref{fig:convergence}, this ensures 2-norm errors below
$10^{-6}$ for resolving the Y-soliton solution up to $t =
100$. Simulations are run on domains of various sizes that depend upon
the problem, typically with an area comparable to $[-512,512]^2$. To
take full advantage of our system's graphics processing unit, we used
single precision for all calculations presented in the main text.

\begin{figure}
  \centering
  \includegraphics[]{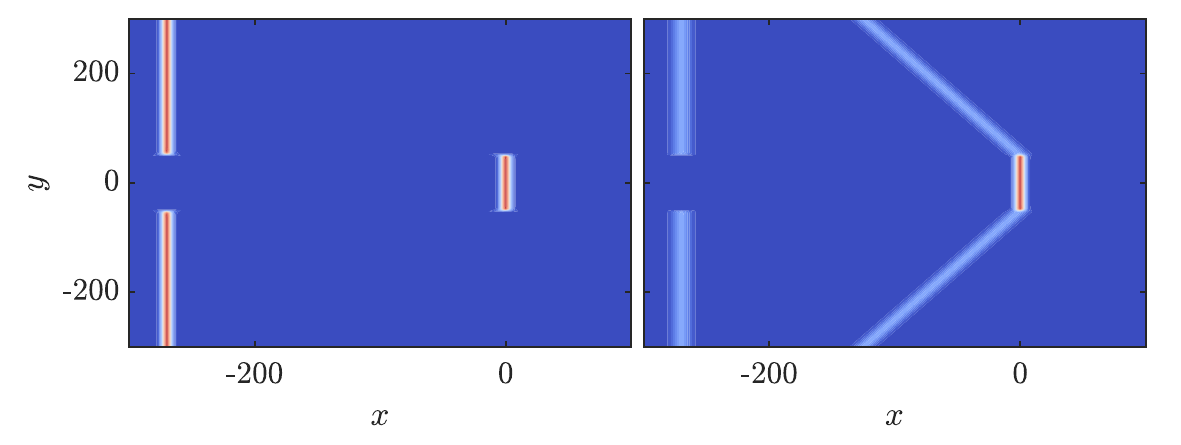}
  \caption{Full initial conditions for simulation of truncated (left)
    and bent-stem solitons (right), satisfying the constraint $\int
    u_{yy}\,\mathrm{d}x = 0$.}
  \label{fig:ref_ic}
\end{figure}

One unique feature of our numerical simulations is the incorporation
of the KP equation constraint requiring $\int u_{yy} dx = 0$ (see
\cite{ablowitz_kp_constraint_1991,Klein_numerical_2007}). The bent
soliton initial conditions satisfy this constraint. For the initial
conditions which do not satisfy the constraint, a reflection is added
that ensures that $\int u_y dx = 0$, thereby satisfying the
constraint. The reflected solitons have parameters $(a_r,q_r)$ defined
by
\begin{equation}
    \label{eq:ref}
  a_r(y) = (1-\sqrt{a(y,0)})^2, \quad q_r = 0.
\end{equation}
The full initial conditions including the reflection for the truncated
and bent-stem cases are displayed in
Fig.~\ref{fig:ref_ic}. Simulations are terminated before the reflected
solitons influence the solution in the region of interest.

To reduce Gibbs phenomenon, initial data for the simulations are
obtained by smoothing the discontinuous initial data $a(y,0)$ and
$q(y,0)$ and inserting this data into Eq.~(\ref{eq:39}).  For example,
the truncated soliton data (\ref{eq:TruncatedIC}) becomes
\begin{equation}
  \label{eq:61}
  u(x,y,0) = a(y,0) \mathrm{sech}^2\left ( \sqrt{\frac{a(y,0)}{12}} x
  \right ), \quad a(y,0) = \frac{1}{2} \left ( \mathrm{tanh}\left ( \frac{\ell/2 - |y|}{w}
    \right ) + 1 \right ) ,
\end{equation}
with $w = 5$, $\ell = 300$.

The two parameters of interest in this paper are the soliton amplitude
and the soliton centre. The amplitude is obtained from the numerical
results by finding the maximum value over $x \in [-L_x,L_x]$ for each
fixed $y$ and $t$ in the domain via local interpolation of the
solution. The soliton centre is simply the location of that maximum
value. These are then compared to the predicted amplitude and centre
values. The predicted soliton centre is calculated using the
analytical solution for $q(y,t)$ combined with \eqref{eq:39} or, if
$y$ is in the simple wave interaction region, using \eqref{eq:s3}. The
numerical values for $q(y,t)$ are found by negating the numerically
differentiated soliton centre location in $x$ with respect to $y$.

\bibliographystyle{plain}


\end{document}